\documentclass[prb,aps,superscriptaddress,twocolumn]{revtex4}
\usepackage{graphicx}
\usepackage{dcolumn}
\usepackage{bm}

\usepackage{epstopdf} 

\usepackage{amsmath, amssymb, graphics, setspace}

\newcommand{\mathsym}[1]{{}}
\newcommand{\unicode}[1]{{}}

\begin{document}

\title{Interplay of many-body and single-particle interactions  in  iridates and rhodates.}

\author{Natalia  B. Perkins}
\affiliation{Department of Physics, University of Wisconsin,
Madison, Wisconsin 53706, USA}
\author{Yuriy Sizyuk}
\affiliation{Department of Physics, University of Wisconsin,
Madison, Wisconsin 53706, USA}

\author{Peter W\"{o}lfle}
\affiliation{Department of Physics, University of Wisconsin,
Madison, Wisconsin 53706, USA}

\affiliation{Institute for Condensed Matter Theory and Institute for Nanotechnology, Karlsruhe Institute of Technology, D-76128 Karlsruhe, Germany}

\begin{abstract}
Motivated by recent experiments exploring the spin-orbit-coupled magnetism in $4d$- and $5d$-band transition metal oxides, we study  magnetic  interactions in  Ir- and Rh-based  compounds. In these systems, the comparable strength of spin-orbit  coupling (SOC), crystal field splitting (CF) and Coulomb and Hund's coupling leads to a rich variety of magnetic exchange interactions, leading to new types of ground states.
Using a strong coupling approach, we derive effective  low-energy super-exchange Hamiltonians from the
multi-orbital Hubbard model by taking full account of the Coulomb and Hund's interactions in the intermediate states.  We find that  in the presence of  strong SOC and lattice distortions the super-exchange Hamiltonian contains  various kinds of magnetic anisotropies.
 Here we are  primarily interested in the  magnetic properties of Sr$_2$IrO$_4$ and Sr$_2$Ir$_{1-x}$Rh$_x$O$_4$ compounds. We perform a systematic study of  how  magnetic interactions in these systems depend on the microscopic parameters and provide a thorough analysis of the resulting magnetic phase diagram. Comparison of our results with experimental data shows good agreement. Finally, we discuss the parameter space in which the spin-flop transition in Sr$_2$IrO$_4$, experimentally observed under pressure, can be realized.
 \end{abstract}

\maketitle

\section{Introduction}

 5$d$ transition metal oxides,  in which
 orbital degeneracy is accompanied by  strong relativistic SOC, recently received considerable attention, both in experiment and in theory.  In these systems the SOC might be comparable to, or even stronger than the
  Coulomb and Hund's couplings, and the CF interactions arising from surrounding oxygen atoms in a nearly octahedral
environment. As a result of this  unusual hierarchy of on-site interactions,
 novel quantum and classical  states with non-trivial topology and interesting magnetic properties might be stabilized.
Fascinating examples of such properties include
the Mott insulator
ground state of Sr$_2$IrO$_4$
\cite{cao98,bjkim08,moon09,kim09,kim12,kimjackeli12,fujiyama12,haskel12,boseggia13,cetin12,jackeli09,fawang11,martins11,min12,arita12},
 the potential spin-liquid
ground state of Na$_4$Ir$_3$O$_8$
\cite{okamoto07,lawler08},
the anomalous Hall effect in the metallic frustrated pyrochlore Pr$_2$Ir$_2$O$_7$\cite{nakatsuji06,machida07,machida10,Pesin10,yang10,lee13},
  non-trivial long-range order, and moment fluctuations in its sister compound Eu$_2$Ir$_2$O$_7$\cite{zhao11,ishikawa12},  unusual magnetic orderings  in the honeycomb compounds
Na$_2$IrO$_3$ and Li$_2$IrO$_3$\cite{shitade09,jackeli10,singh10,singh12,liu11,ye12,price12,price13},
and others.

\begin{figure*}\label{geometry}
\includegraphics[width=0.69\columnwidth]{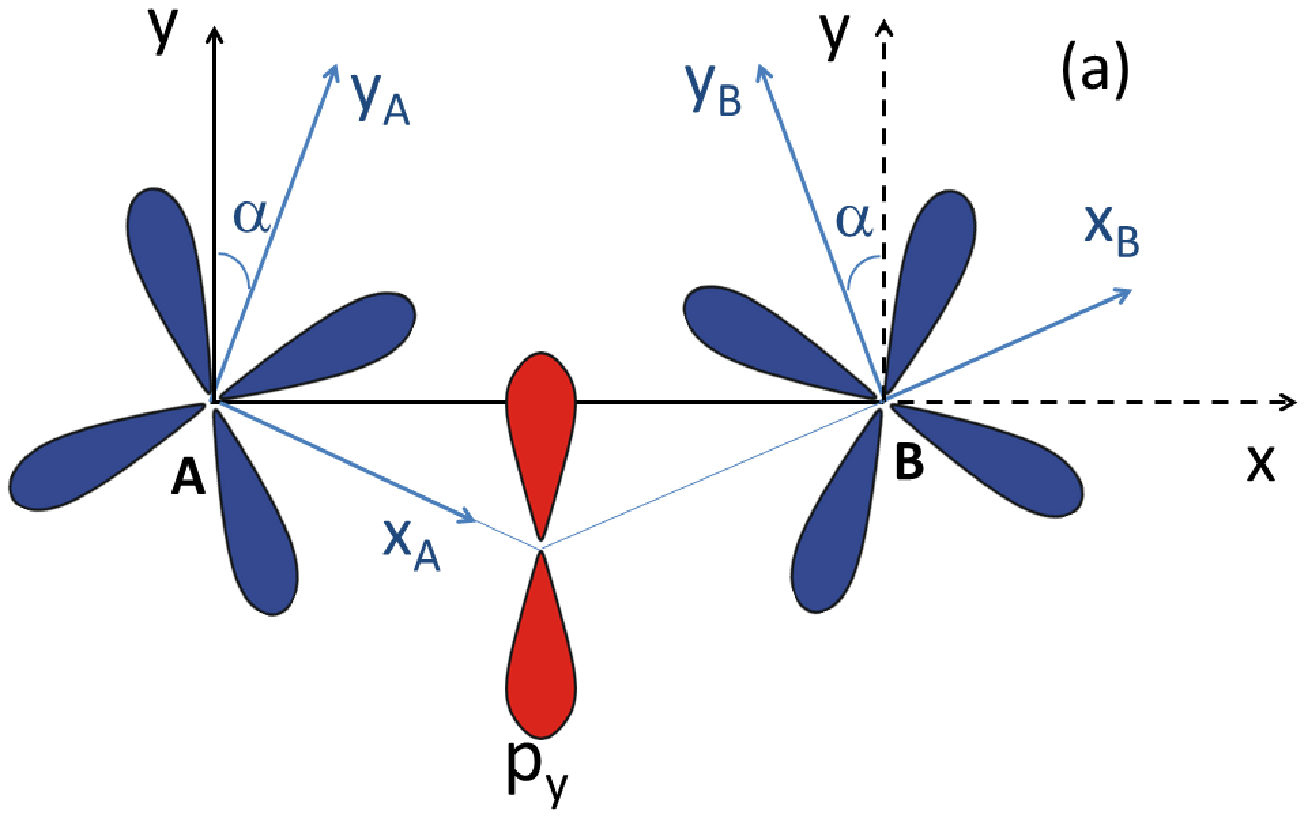}
\includegraphics[width=0.72\columnwidth]{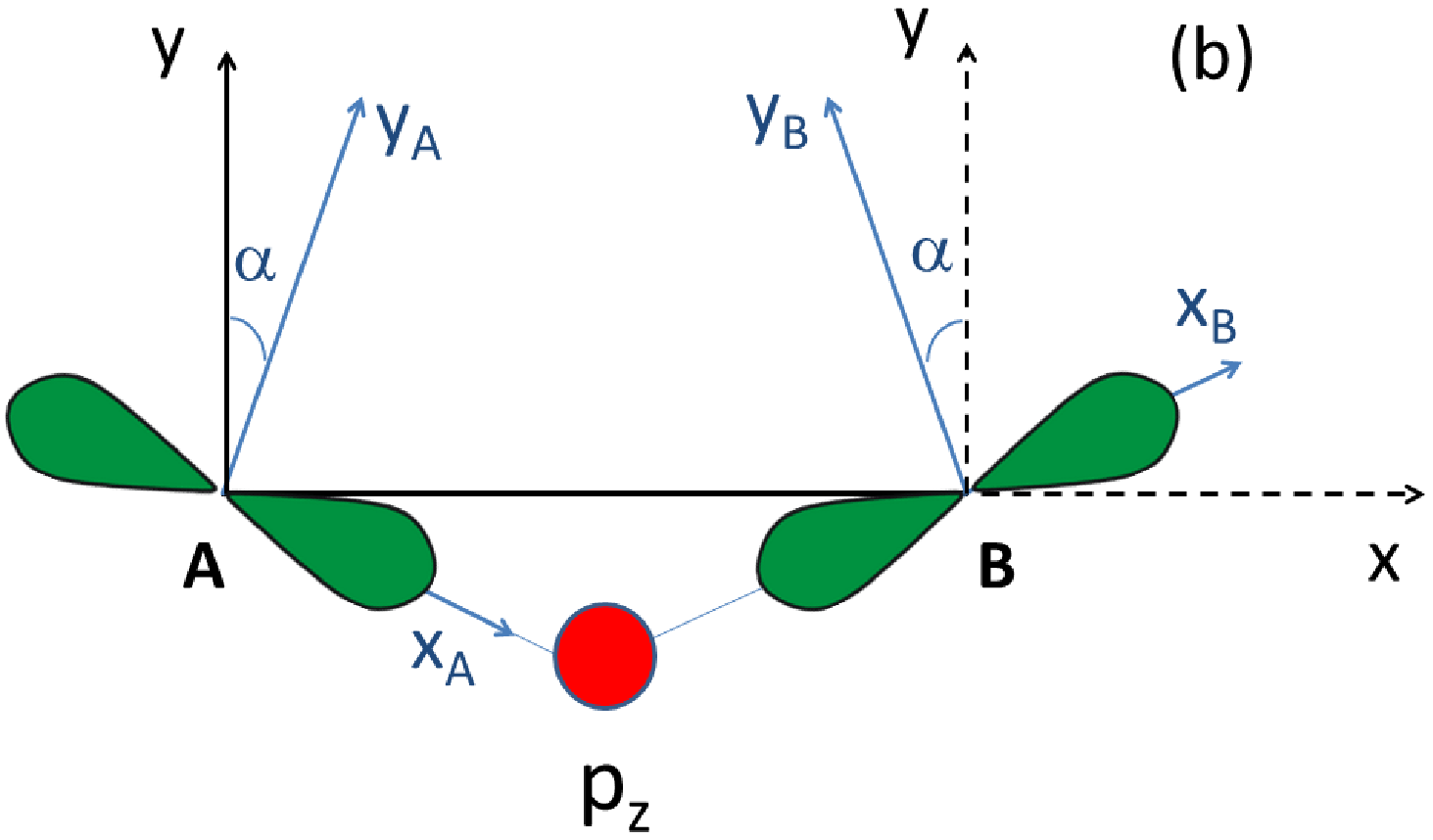}
\caption{ Ir-O-Ir bond  in the presence of
octahedra rotations in one IrO$_2$ layer.
$x$ and $y$ are the global axes adopted for the intermediate oxygen atoms.
 $x_{A(B)}$ and $y_{A(B)}$ are local axes on sublattices A and B. (a): Local ${\tilde Z}$ orbital on Ir ion overlaps with $p_y$ oxygen orbital in the global reference frame. (b): Local ${\tilde Y}$ orbital on Ir ion overlaps with $p_z$ oxygen orbital in the global reference frame.}
\end{figure*}

The  main focus of this paper is on developing a theoretical framework  which will allow us to understand  the microscopic nature of magnetism in the  iridium compounds  described above. In  these systems, the magnetic degrees of freedom are determined by  Ir$^{4+}$ ions  in $5d^5$  electronic configurations.
 In contrast to the 3$d$-based oxides, in which a Mott insulating state is established by  strong correlations,
 in iridates with  Ir$^{4+}$ ions, the Mott insulating state  does not occur without sufficiently strong SOC.~\cite{bjkim08} This is because the relatively small Coulomb interaction is too weak to open a charge gap for these systems with broad $t_{2g}$ bands at a 5/6 filling factor, such that the systems would be either metals or band insulators.  The SOC, however, splits the $t_{2g}$ bands into  ${\mathbf J} = 1/2$ and ${\mathbf J} = 3/2$ bands. Four  out of five electrons fill the lower ${\mathbf J}= 3/2$ band, and one electron half-fills the ${\mathbf J} = 1/2$ band. Because the ${\mathbf J} = 1/2$ band is relatively narrow, the Coulomb repulsion is sufficient to open a  charge gap in the ${\mathbf J} = 1/2$ band   and, therefore, a Mott insulating state occurs. In an idealized system without lattice distortions the magnetic degrees of freedom of this insulating state can be described by the ${\mathbf J} = 1/2$ Kramers  doublet.

The super-exchange Hamiltonians  for layered iridium oxides were first derived in the seminal paper by Jackeli and Khalliulin.~\cite{jackeli09}  They found that the super-exchange  Hamiltonian describing the coupling between ${\mathbf J} = 1/2$ Kramers  doublet states  on the square lattice, like in Sr$_2$IrO$_4$,  is predominantly  of isotropic Heisenberg super-exchange character,  while anisotropic terms become relevant only in the presence of lattice distortions.
 On the honeycomb lattice,  like in Na$_2$IrO$_3$, the interaction between ${\mathbf J} = 1/2$ Kramers  doublet states is  highly anisotropic  even in the absence of lattice distortions.  The anisotropic part of the super-exchange coupling has the very peculiar form of the Kitaev interaction. This originates  from the competition between SOC and correlation effects, and is non-zero only in the presence of Hund's coupling.

In the present study, we go beyond this  work and derive effective super-exchange spin Hamiltonians rigorously starting from the exact eigenstates of the single ion  microscopic  Hamiltonian. Here, we will be primarily interested in the magnetic properties of the insulating iridium oxides with tetragonal symmetry,
 in which the  Ir ions occupy a square  lattice, as in the case of Sr$_2$IrO$_4$. While this particular compound  is very interesting and has recently attracted much attention,~\cite{cao98,bjkim08,moon09,kim09,kim12,kimjackeli12,fujiyama12,haskel12,cetin12,jackeli09,fawang11,martins11,min12,arita12} the approach proposed here can not only be easily used to understand the magnetic properties of other iridates belonging to the Ruddlesden-Popper series Sr$_{n+1}$Ir$_n$O$_{3n+1}$, but can also be applied to systems with different lattice geometries.

The magnetic properties of Sr$_2$IrO$_4$ are very unusual.  Below 240 K, Sr$_2$IrO$_4$  is a canted antiferromagnet  with a small in-plane ferromagnetic  moment ($\sim 0.1\mu_B$)~\cite{cao98}, which, however, is one to two orders of magnitude
larger than that of the analogous canted antiferromagnet La$_2$CuO$_4$.
Another important experimental finding is that this canting disappears with pressure.~\cite{haskel12}
These two observations
indicate a very strong coupling between magnetic properties  and the crystal  lattice, which in the presence of  SOC   can be understood through  the coupling of orbital magnetization to the lattice. Consequently, as the orbital  magnetic moment contributes to the total magnetic moment of Ir ions, there is a strong dependence of the magnetic degrees on lattice degrees of freedom.

Two types of  lattice distortions are present in Sr$_2$IrO$_4$ even at ambient pressure: tetragonal distortion and staggered rotation of IrO$_6$ octahedra (see Fig.1). The staggered nature of the IrO$_6$ octahedra rotation leads to a doubling of the unit cell and the formation of a two-sublattice structure.
 The tetragonal distortion moves the electronic ground state away from the strong SO limit  ${\mathbf J} = 1/2$ state by mixing ${\mathbf J} = 1/2$  and ${\mathbf J} = 3/2$ states. Thus in order to understand the magnetism of this system, one needs  first to understand the nature of the magnetic degrees of freedom.  In our approach we identify the magnetic degrees of freedom by dealing  with  the exact eigenstates of the full single ion  microscopic  Hamiltonian which includes both SOC and CF interactions.

In this work, we will obtain dependencies of the magnetic interactions on microscopic parameters characterizing the system. In addition, we will  study  how the properties of Sr$_2$IrO$_4$  depend on  external pressure\cite{haskel12} and chemical substitution.~\cite{boseggia13,qi12,luo13} In particular, we will discuss   the case when iridium is substituted with rhodium.\cite{qi12,luo13}
 Rh substitution, unlike other chemical substitutions, does not change
the band filling.  However, it  varies  the SOC, and the Coulomb  and Hund's coupling strengths, because  on one side the 4$d$ orbitals of Rh ions are less extended, tending to enhance the electronic repulsion and thereby increasing correlation effects, and on the other side, as  Rh is a lighter ion, the SOC  is smaller.
   Thus, when the content of Rh increases, the overall balance of on-site interactions changes, and as a result one might expect  the appearance of new magnetic phases and  doping-driven  phase transitions.     Although this direction has been recently explored experimentally in a few cases~\cite{qi12,luo13},
 it still remains to be investigated theoretically.

The paper is organized as follows. In Sec.~\ref{sec:model}, we  introduce
the single ion microscopic model appropriate for the description of the physical
properties of the  iridates and rhodates, in which five electrons  or, equivalently, one hole occupy the three-fold degenerate $t_{2g}$ orbitals and experience strong SOC and crystal field (CF) interactions. We  first obtain one-particle eigenstates  taking into account only SOC and CF interactions, and then  compute two-particle excited eigenstates   fully considering  correlation effects. In Sec.~\ref{sec:ham} we derive an effective super-exchange Hamiltonian by  integrating out the intermediate oxygen ions  and performing a second order perturbation expansion in the hopping parameters around the
atomic limit.
   In Sec.~\ref{sec:results}, we  present our results on the magnetic interactions and show how  these interactions depend on various microscopic parameters of the model.
   We also  discuss the application of the results  obtained to real compounds.
     Finally, in Sec.~\ref{sec:conclusion} a summary of the work is  presented.

\section{Single ion Hamiltonian}\label{sec:model}

\subsection{One-particle eigenstates}

  In Sr$_2$IrO$_4$, the Ir$^{+4}$ ions are   sitting inside an oxygen  cage forming an octahedron. The octahedral crystal field splits the five 5$d$ orbitals of Ir into a doublet of $e_g$
orbitals at higher energy  and into the low-lying three-fold degenerate $t_{2g}$ multiplet.
  In iridates, the energy difference between the $e_g$ and  $t_{2g}$ levels is large. Because of this, the five electrons occupy  the low lying $t_{2g}$ orbitals and the on-site interactions, such as the SOC, Coulomb  and Hund's interactions, and the crystal field interactions, lowering the symmetry further,  can be considered within the $t_{2g}$ manifold only. In this limit, the  SOC has to be projected to the $t_{2g}$  manifold, resulting in an
effective orbital angular momentum $L=1$.

It is more convenient to describe the low spin state of the
$d^5$-configuration of Ir$^{+4}$ ions by using the hole description.
In the local axes bound to the oxygen octahedron the $t_{2g}$ orbitals  of Ir ions are
$\left\vert X\right\rangle =\left\vert yz\right\rangle $,
$\left\vert Y\right\rangle =\left\vert zx\right\rangle $, and
$\left\vert Z\right\rangle=\left\vert xy\right\rangle $. Examples of the lobe structure of the d-wave orbital $\left\vert xy\right\rangle $ of Ir are shown in Fig.1  (blue lobes).
In the absence of interactions, these one-hole states are completely degenerate.
 The SOC  and CF  interactions, described by the single-ion Hamiltonian
\begin{eqnarray}\label{ham_CF-SOC}
H_{\lambda,\Delta}=\lambda \overrightarrow{S}\cdot \overrightarrow{L}
+\Delta L_{z}^{2},
\end{eqnarray}
give rise to a splitting of the levels according to the symmetry of the underlying lattice.
In the tetragonal system, the orbital angular momentum basis is defined by
$\vert L_{z}=0\rangle =\vert Z\rangle $,
 $ \vert L_{z}=\pm 1\rangle =-\frac{1}{\sqrt{2}}(\pm\vert X\rangle +
\imath \vert Y\rangle )$, where the quantization axis is taken along the tetragonal $z$ axis.
In the absence of tetragonal distortion, the energy eigenstates are the angular momentum eigenstates $\vert
J,J_{z}\rangle$. The full single-particle Hilbert space is, thus, given   by a six-component vector
$
{\hat J}=\{
|\frac{1}{2},\frac{1}{2}\rangle,
|\frac{1}{2},-\frac{1}{2}\rangle,
|\frac{3}{2},\frac{3}{2}\rangle,
|\frac{3}{2},\frac{1}{2}\rangle ,
|\frac{3}{2},-\frac{1}{2}\rangle,
|\frac{3}{2},\frac{-3}{2}\rangle\}
$.
The vector ${\hat J}$ can be expressed in terms of the basis set of $t_{2g}$-orbitals
as
\begin{eqnarray}\label{jbasistoxbasis}
 {\hat J} =
 \left( \begin{array}{cccccc}
0 & -\frac{1}{\sqrt{3}} & 0 & -\frac{\imath}{\sqrt{3}} & -\frac{1}{\sqrt{3}} & 0 \\
-\frac{1}{\sqrt{3}} & 0 & \frac{\imath}{\sqrt{3}} & 0 & 0 & \frac{1}{\sqrt{3}}\\
-\frac{1}{\sqrt{2}} & 0 & -\frac{\imath}{\sqrt{2}} & 0 & 0 & 0 \\
0 & -\frac{1}{\sqrt{6}} & 0 & -\frac{\imath}{\sqrt{6}} &\sqrt{\frac{2}{3}} & 0 \\
\frac{1}{\sqrt{6}} & 0 & -\frac{\imath}{\sqrt{6}} & 0 & 0 & \sqrt{\frac{2}{3}}  \\
0 & \frac{1}{\sqrt{2}} & 0 & -\frac{\imath}{\sqrt{2}} & 0 & 0
\end{array} \right ) {\hat A}_1~,
\end{eqnarray}
where
${\hat A}_1=\{
\vert X_\uparrow\rangle,
\vert X_\downarrow\rangle,
\vert Y_\uparrow\rangle,
\vert Y_\downarrow\rangle,
\vert Z_\uparrow\rangle,
\vert Z_\downarrow\rangle \}$ is a six-component vector, and $\uparrow,\downarrow$ indicate spin states.
The ground state is  a Kramers doublet $\vert J=\frac{1}{2},J_{z}\rangle$ at  energy $E_{0}=-\lambda $  and the
excited state form a quartet $\vert J=\frac{3}{2},J_{z}\rangle $ at energy $E_{1}=\frac{1}{2}\lambda$.
However, in Sr$_2$IrO$_4$, the tetragonal distortion is  present  and is not small. It  arises because the oxygen octahedra are elongated along the $z$ axis. In the hole representation, $\Delta=\Delta_{tet}  >0$ and the  $t_{2g}$ orbitals are split into a singlet state $\vert Z\rangle$ with energy $-\Delta$ and a doublet state  ($\vert X\rangle \& \vert Y\rangle$) with energy $\Delta/2$.
In the presence of both the tetragonal distortion and the SOC,
the eigenfunctions of the Hamiltonian (\ref{ham_CF-SOC})  are given by  components of
a vector ${\hat \Psi}=\{
|\Psi_1\rangle,
|\Psi_2\rangle,
|\Psi_3\rangle,
|\Psi_4\rangle,
|\Psi_5\rangle,
|\Psi_6\rangle \}$, which in terms of $t_{2g}$-orbitals are given by
\begin{eqnarray}\label{psi}
{\hat \Psi}={\hat M}^{tet}_{\theta}{\hat A}_1~,
\end{eqnarray}
where
\begin{eqnarray} \label{psitoxbasis}\nonumber 
{\hat M}^{tet}_{\theta}
=
 \left( \begin{array}{cccccc}
0&\frac{1}{\sqrt{2}}\,c_{\theta}&0&\frac{\imath }{\sqrt{2}}\,c_{\theta}&s_{\theta}&0\\
-\frac{1}{\sqrt{2}}\,c_{\theta}&0&\frac{\imath }{\sqrt{2}}\,c_{\theta}&0&0&s_{\theta}\\
-\frac{1}{\sqrt{2}}&0&-\frac{\imath}{\sqrt{2}}&0&0&0\\
0&-\frac{1}{\sqrt{2}}\,s_{\theta}&0&-\frac{\imath }{\sqrt{2}}\,s_{\theta}&c_{\theta}&0\\
\frac{1}{\sqrt{2}}\,s_{\theta}&0&-\frac{\imath}{\sqrt{2}}\,s_{\theta}&0&0&c_{\theta}\\
0&-\frac{1}{\sqrt{2}}&0&\frac{\imath}{\sqrt{2}}&0&0
\end{array} \right ),
\end{eqnarray}
where, for shortness, we denote $c_{\theta}=\cos\theta$ and $s_{\theta}=\sin\theta$.
The angle variable $\theta$ is determined by $\tan (2\theta )=2\sqrt{2}\frac{\lambda }{\lambda -2\Delta }$ and takes care of the competition between the tetragonal distortion and the SOC.\cite{jackeli09}

The eigenstates  of the Hamiltonian (\ref{ham_CF-SOC})  are given by the following three doublets:
the  ground state doublet ($\vert\Psi_1\rangle \& \vert\Psi_2\rangle$)
 with energy
$E^{(1,2)}=\frac{1}{2}(\Delta -\frac{\lambda }{2})-\frac{1}{2}\sqrt{%
2\lambda ^{2}+(\Delta -\frac{\lambda }{2})^{2}}=-\frac{\lambda }{\sqrt{2}}\cot \theta $, the intermediate doublet
($\vert\Psi_4\rangle \& \vert\Psi_5\rangle$)
 with energy
$E^{(4,5)}=\frac{1}{2}(\Delta -\frac{\lambda }{2})+\frac{1}{2}\sqrt{%
2\lambda ^{2}+(\Delta -\frac{\lambda }{2})^{2}}=\frac{\lambda }{\sqrt{2}}\tan \theta$
and  the  upper doublet ($\vert\Psi_3\rangle \& \vert\Psi_6\rangle$)
 with energy
 $E^{(3,6)}=\Delta +\frac{\lambda }{2}$ .
Note that the ground state doublet $\vert\Psi_1\rangle \& \vert\Psi_2\rangle$ is different from the $|J=\frac{1}{2},J_z\rangle$ doublet as
well as the $L_{z}=0$ doublet!

In Sr$_2$IrO$_4$,  there is also a staggered rotation of  neighboring oxygen octahedra  by an angle $\pm \alpha$ about the $z$-axis (see Fig.1 (a) and (b)) leading to the formation  of a two-sublattice structure. We denote these two sublattices as A and B.
 Because the crystal-field interaction on Ir 5$d$ orbitals
 is diagonal only in the local cubic axes bound to the oxygen octahedron, in the presence  of the octahedra rotations, atomic states on sublattices A  and B have to be defined in the local  basis. Then, the states on  sublattices  A and B are given by
\begin{eqnarray} \label{psitoxbasisA}
{\hat\Psi}_A
={\hat M}^{tet}_{\theta}
 \left( \begin{array}{l}
\vert {\tilde X}_\uparrow e^{-\frac{\imath\alpha}{2}}\rangle  \\
\vert {\tilde X}_\downarrow e^{\frac{\imath\alpha}{2}}\rangle  \\
\vert {\tilde Y}_\uparrow e^{-\frac{\imath\alpha}{2}}\rangle \\
\vert {\tilde Y}_\downarrow e^{\frac{\imath\alpha}{2}} \rangle \\
\vert {\tilde Z}_\uparrow e^{-\frac{\imath\alpha}{2}}\rangle  \\
\vert {\tilde Z}_\downarrow e^{\frac{\imath\alpha}{2}}\rangle
 \end{array} \right )
\end{eqnarray}
 and
 \begin{eqnarray} \label{psitoxbasisB}
{\hat\Psi}_B
={\hat M}^{tet}_{\theta}
 \left( \begin{array}{l}
\vert {\tilde X}_\uparrow e^{\frac{\imath\alpha}{2}}\rangle  \\
\vert {\tilde X}_\downarrow e^{-\frac{\imath\alpha}{2}}\rangle  \\
\vert {\tilde Y}_\uparrow e^{\frac{\imath\alpha}{2}}\rangle \\
\vert {\tilde Y}_\downarrow e^{-\frac{\imath\alpha}{2}} \rangle \\
\vert {\tilde Z}_\uparrow e^{\frac{\imath\alpha}{2}}\rangle  \\
\vert {\tilde Z}_\downarrow e^{-\frac{\imath\alpha}{2}}\rangle
 \end{array} \right ),
\end{eqnarray}
where the phase factors $e^{\pm\frac{\imath\alpha}{2}}$  appear after the projection of the spin states onto the local reference frame. Initially the spin states are defined in the global reference frame.

\subsection{Two-hole states in the presence of interactions, spin-orbit coupling and tetragonal distortion.}
The many-body part of the single ion Hamiltonian  is given by
 the three-band  Hubbard Hamiltonian
of the form:
\begin{eqnarray}
H_{int}&=&U_{1}\sum_{i,\alpha}n_{i\alpha\uparrow}n_{i\alpha\downarrow}+
\frac{1}{2}(U_2-J_H)\sum_{i,\sigma, \alpha\neq \alpha '}
n_{i\alpha\sigma}n_{i\alpha'\sigma}  \nonumber \\
&+&U_2\sum_{i,\alpha\neq \alpha'}n_{i\alpha\uparrow}n_{i\alpha'\downarrow}+
J_H\sum_{i,\alpha\neq \alpha'} d_{i\alpha\uparrow}^\dagger
d_{i\alpha\downarrow}^\dagger d_{i\alpha'\downarrow}
d_{i\alpha'\uparrow} \nonumber \\
&-&J_H\sum_{i,\alpha\neq \alpha'} d_{i\alpha\uparrow}^\dagger
d_{i\alpha\downarrow}d_{i\alpha'\downarrow}^\dagger
d_{i\alpha'\uparrow} ~,
\label{h2}
\end{eqnarray}
\noindent where $U_1$ and $U_2$ are the Coulomb repulsion among electrons in the same and in different $t_{2g}$ orbitals, respectively, and
$J_H$ is the  Hund's coupling constant. Due to the cubic symmetry, the relation
$U_1=U_2+2J_H$ holds. The annihilation and creation electron operators,
$d_{i\alpha\sigma}$ and $d^{\dagger}_{i\alpha\sigma}$
refer to Ir orbitals at site $i$, of
type $\alpha$ ($X$, $Y$ or $Z$)
and with spin $\sigma=\uparrow,\downarrow$,
$n_{i\alpha\sigma}=d^{\dagger}_{i\alpha\sigma}d_{i\alpha\sigma}$.
In  order to obtain $H_{int}$ in the hole picture,  we substitute
$d_{\alpha \sigma }^{\dag }\rightarrow a_{\alpha \sigma }$  and  $%
n_{\alpha \sigma }\rightarrow 1-h_{\alpha \sigma }$, where $%
h_{\alpha \sigma }=a_{\alpha \sigma }^{\dag }a_{\alpha \sigma }$, and $a_{\alpha \sigma }^{\dag }$ and $a_{\alpha \sigma }$ are the hole creation and annihilation operators.

We first compute energy eigenvalues of $H_{int}$. We consider ground states with one hole on the Ir ion and
excited states, in which the  Ir ion can have  two holes or no holes.
The corresponding energies are
\begin{eqnarray}\begin{array}{l}
E_{1h}=10U_2\\[0.1cm]
E_{0h}=15U_2\\[0.1cm]
E^{(1)}_{2h}=6U_2-J_H\\[0.1cm]
E^{(0)}_{2h}=6U_2+J_H\\[0.1cm]
E^{(00)}_{2h}=6U_2+4J_H
\end{array}
\label{exener}
\end{eqnarray}
There are $6\times 5/2=15$ partly degenerate two-hole states: six spin
singlets and three triplets. Let the vector $\vert {\mathcal I}\rangle=\vert {\mathcal I};n\rangle $ denote the two-hole eigenstates.
It is convenient to represent them using the cubic orbital basis:
\begin{eqnarray}
\hat{\mathcal I}= \hat{{\mathcal M}}_2\hat{A}_2,
\end{eqnarray}
 where
\begin{eqnarray}
{\hat A}_2= &\{&X_\uparrow X_\downarrow,X_\uparrow Y_\uparrow,X_\uparrow Y_\downarrow,X_\uparrow Z_\uparrow,X_\uparrow Z_\downarrow,
   X_\downarrow Y_\uparrow,X_\downarrow Y_\downarrow,\nonumber\\\nonumber
   &&X_\downarrow Z_\uparrow,X_\downarrow Z_\downarrow,
   Y_\uparrow Y_\downarrow,Y_\uparrow Z_\uparrow, Y_\uparrow Z_\downarrow,Y_\downarrow Z_\uparrow, Y_\downarrow Z_\downarrow,
   Z_\uparrow Z_\downarrow \}
   \end{eqnarray}
 is the two-hole orbital basis and  the transformation matrix $\hat{{\mathcal M}}_2$  can be easily obtained. Explicitly, vector  $\hat{\mathcal I}$ consists of the following elements:\\
(i) symmetric state with singlet pairs on the same orbital $S=0,\alpha =\alpha ^{\prime }$
\begin{eqnarray}\nonumber
\vert {\mathcal I};1\rangle  =%
\frac{1}{\sqrt{3}}\left(a_{X\downarrow }^\dagger a_{X\uparrow }^\dagger+a_{Y\downarrow
}^\dagger a_{Y\uparrow }^\dagger +a_{Z\downarrow }^\dagger a_{Z\uparrow }^\dagger \right)\vert vac\rangle
\end{eqnarray}
with  energy equal to $E_{1}=E_{2h}^{(00)}=6U_{2}+4J_{H}=E_{d}$.\\
(ii) two degenerate antisymmetric states with singlet pairs on the same orbital $S=0,\alpha =\alpha ^{\prime }$
\begin{eqnarray}
\vert {\mathcal I};2\rangle
&=&\frac{1}{\sqrt{2}}\left(a_{X\downarrow }^\dagger a_{X\uparrow }^\dagger-a_{Y\downarrow
}^\dagger a_{Y\uparrow }^\dagger \right)\vert vac\rangle\nonumber\\\nonumber
\vert {\mathcal I};3\rangle
&=&\frac{1}{\sqrt{6}}\left(a_{X\downarrow }^\dagger a_{X\uparrow }^\dagger +a_{Y\downarrow
}^\dagger a_{Y\uparrow }^\dagger -2a_{Z\downarrow }^\dagger a_{Z\uparrow }^\dagger \right)\vert vac\rangle
\end{eqnarray}
with  energies equal to
$E_{2,3}=E_{2h}^{(0)}=6U_{2}+J_{H}=E_{s}$.\\
(iii) three states with singlet pairs on different orbitals $S=0,\alpha \neq \alpha ^{\prime }$:\\
\begin{eqnarray}
\vert {\mathcal I};4\rangle  &=&\frac{1}{
\sqrt{2}}\left( a_{X\downarrow }^\dagger a_{Y\uparrow }^\dagger-a_{X\uparrow
}^\dagger a_{Y\downarrow }^\dagger \right)\vert vac\rangle\nonumber\\
\vert {\mathcal I};5\rangle &=&\frac{1}{
\sqrt{2}}\left( a_{Y\downarrow }^\dagger a_{Z\uparrow }^\dagger-a_{Z\uparrow
}^\dagger a_{Y\downarrow }^\dagger \right)\vert vac\rangle \nonumber\\
\vert {\mathcal I};6\rangle &=&\frac{1}{
\sqrt{2}}\left( a_{Z\downarrow }^\dagger a_{X\uparrow }^\dagger-a_{X\uparrow
}^\dagger a_{Z\downarrow }^\dagger \right)\vert vac\rangle \nonumber
\end{eqnarray}
with $E_{4,5,6}=E_{2h}^{(0)}=6U_{2}+J_{H}=E_{s}$.\\
(iv) Nine states with triplet pairs on different orbitals $S=1,\alpha \neq \alpha ^{\prime }$:
\begin{eqnarray}
&&\vert {\mathcal I};7\rangle =
\frac{1}{\sqrt{2}}\left(a_{X\downarrow }^\dagger a_{Y\uparrow }^\dagger +a_{X\uparrow
}^\dagger a_{Y\downarrow }^\dagger \right)\vert vac\rangle\nonumber\\
&&\vert {\mathcal I};8\rangle
= a_{X\uparrow }^\dagger a_{Y\uparrow }^\dagger\vert vac\rangle \nonumber\\
&&\vert {\mathcal I};9\rangle
=a_{X\downarrow }^\dagger a_{Y\downarrow }^\dagger\vert vac\rangle \nonumber\\
&&\vert {\mathcal I};10\rangle
=
\frac{1}{\sqrt{2}}\left(a_{Y\downarrow }^\dagger a_{Z\uparrow }^\dagger +a_{Y\uparrow
}^\dagger a_{Z\downarrow }^\dagger \right)\vert vac\rangle\nonumber\\
&&\vert {\mathcal I};11\rangle
= a_{Y\uparrow }^\dagger a_{Z\uparrow }^\dagger\vert vac\rangle \nonumber\\
&&\vert {\mathcal I};12\rangle
=a_{Y\downarrow }^\dagger a_{Z\downarrow }^\dagger\vert vac\rangle \nonumber\\
&&\vert {\mathcal I};13\rangle =
\frac{1}{\sqrt{2}}\left(a_{Z\downarrow }^\dagger a_{X\uparrow }^\dagger +a_{Z\uparrow
}^\dagger a_{X\downarrow }^\dagger \right)\vert vac\rangle\nonumber\\
&&\vert {\mathcal I};14\rangle
= a_{Z\uparrow }^\dagger a_{X\uparrow }^\dagger\vert vac\rangle \nonumber\\
&&\vert {\mathcal I};15\rangle
=a_{Z\downarrow }^\dagger a_{X\downarrow }^\dagger\vert vac\rangle \nonumber
\end{eqnarray}
with energies $E_{7,..,15}=E_{4}^{(1)}=6U_{2}-J_{H}=E_{t}$.
This gives three different excitation energies:
\begin{eqnarray}
\Delta E_{1}&=&E_d+E_{0h}-2E_{1h}=U_{2}+4J_{H}\nonumber\\
\Delta E_{2}&=&E_s+E_{0h}-2E_{1h}=U_{2}+J_{H}\\
\Delta E_{3}&=&E_t+E_{0h}-2E_{1h}=U_{2}-J_{H}\nonumber
\end{eqnarray}

In the presence of the SOC and lattice distortions, the two-hole states  $\vert {\mathcal I};n\rangle$ are mixed, and the true two-hole eigenstates are obtained by diagonalization of the full on-site Hamiltonian
 \begin{eqnarray}\label{ham}
 H_{int+\lambda,\Delta}=H_{int}+H_{\lambda,\Delta}~.
 \end{eqnarray}
  To this end, it is convenient first to  represent the $\vert {\mathcal I};n\rangle$ states in terms of the eigenstates of $H_{\lambda,\Delta}$.
 The two-hole eigenstates of the  SOC part of the  Hamiltonian are simply given by  product states $\vert {\mathcal J},\mu\rangle \equiv\vert J_{1},J_{1z};J_{2},J_{2z}\rangle$:
\begin{eqnarray}\label{jjzjjz}
\begin{array}{lll}
\vert {\mathcal J},1\rangle &\equiv& \vert \frac{1}{2},\frac{1}{2};\frac{3}{2},\frac{3}{2}\rangle\\[0.1cm]
\vert {\mathcal J},2\rangle &\equiv &\vert \frac{1}{2},-\frac{1}{2};\frac{3}{2},\frac{3}{2}\rangle\\[0.1cm]
\vert {\mathcal J},3\rangle &\equiv &\vert \frac{1}{2},\frac{1}{2};\frac{3}{2},\frac{1}{2}\rangle\\[0.1cm]
\vert {\mathcal J},4\rangle &\equiv &\vert \frac{1}{2},-\frac{1}{2};\frac{3}{2},\frac{1}{2}\rangle\\[0.1cm]
\vert {\mathcal J},5\rangle &\equiv &\vert \frac{1}{2},\frac{1}{2};\frac{3}{2},-\frac{1}{2}\rangle\\[0.1cm]
\vert {\mathcal J},6\rangle &\equiv &\vert \frac{1}{2},-\frac{1}{2};\frac{3}{2},-\frac{1}{2}\rangle\\[0.1cm]
\vert {\mathcal J},7\rangle &\equiv &\vert \frac{1}{2},\frac{1}{2};\frac{3}{2},-\frac{3}{2}\rangle\\[0.1cm]
\vert {\mathcal J},8\rangle &\equiv &\vert \frac{1}{2},-\frac{1}{2};\frac{3}{2},-\frac{3}{2}\rangle\\[0.1cm]
\vert {\mathcal J},9\rangle &\equiv &\vert \frac{1}{2},\frac{1}{2};\frac{1}{2},-\frac{1}{2}\rangle\\[0.1cm]
\vert {\mathcal J},10\rangle &\equiv &\vert \frac{3}{2},\frac{1}{2};\frac{3}{2},\frac{3}{2}\rangle\\[0.1cm]
\vert {\mathcal J},11\rangle &\equiv &\vert \frac{3}{2},-\frac{1}{2};\frac{3}{2},\frac{3}{2}\rangle\\[0.1cm]
\vert {\mathcal J},12\rangle &\equiv &\vert \frac{3}{2},-\frac{3}{2};\frac{3}{2},\frac{3}{2}\rangle\\[0.1cm]
\vert {\mathcal J},13\rangle &\equiv &\vert \frac{3}{2},-\frac{1}{2};\frac{3}{2},\frac{1}{2}\rangle\\[0.1cm]
\vert {\mathcal J},14\rangle &\equiv &\vert \frac{3}{2},-\frac{3}{2};\frac{3}{2},\frac{1}{2}\rangle\\[0.1cm]
\vert {\mathcal J},15\rangle &\equiv &\vert \frac{3}{2},-\frac{3}{2};\frac{3}{2},-\frac{1}{2}\rangle
\end{array}
\end{eqnarray}
In short, these states can be written as
\begin{eqnarray}
\vert {\mathcal J},\mu \rangle =\sum_{\gamma _{1},\gamma
_{2}=1}^{6}m_{\gamma _{1},\gamma _{2}}^{\mu }b_{\gamma
_{1}}^\dagger b_{\gamma _{2}}^\dagger \vert vac \rangle,
 \end{eqnarray}
 where $\mu=1,...,15$ refers  to  the component of the  vector ${\hat {\mathcal J}}$,
$b_{\gamma}^\dagger $ is an operator creating a hole  of the type $\gamma=1,...6$, which  refers to  the component of the single-hole vector ${\hat J}$. The tensor ${\hat m}$ has
the following non-zero elements
\begin{eqnarray}\nonumber
m_{1,3}^{1}&=&m_{2,3}^{2}=m_{1,4}^{3}=m_{2,4}^{4}
=m_{1,5}^{5}=\\\nonumber
m_{2,5}^{6}&=&m_{1,6}^{7}=m_{2,6}^{8}=m_{1,2}^{9}
=m_{4,3}^{10}=\\\nonumber
m_{5,3}^{11}&=&m_{6,3}^{12}=m_{5,4}^{13}=m_{6,4}^{14}=m_{6,5}^{15}=1~.
\end{eqnarray}
If, in addition to the SOC, the lattice distortion is present, the two-hole states $\vert {\tilde {\mathcal J}},\mu\rangle$ are  given by the products of two $\vert\Psi_n\rangle$  states.  The explicit form of the vector $\vert {\tilde{\mathcal J}},\mu\rangle $  can be easily obtained from Eq.(\ref{jjzjjz}) by the following substitution:
 \begin{eqnarray}
 \begin{array}{lll}
 \vert \frac{1}{2},\frac{1}{2}\rangle &\rightarrow & \vert \Psi_1 \rangle ,\\[0.1cm]
\vert \frac{1}{2},-\frac{1}{2}\rangle &\rightarrow& \vert \Psi_2 \rangle ,\\[0.1cm]
\vert \frac{3}{2},\frac{3}{2}\rangle &\rightarrow& \vert \Psi_3 \rangle ,\\[0.1cm]
\vert \frac{3}{2},\frac{1}{2}\rangle &\rightarrow& \vert \Psi_4\rangle ,\\[0.1cm]
\vert \frac{3}{2},-\frac{1}{2}\rangle &\rightarrow& \vert \Psi_5 \rangle ,\\[0.1cm]
\vert \frac{3}{2},-\frac{3}{2}\rangle &\rightarrow& \vert \Psi_6 \rangle .
\end{array}
\end{eqnarray}
 The complete Hamiltonian matrix has the same block diagonal structure  in the space of states $%
\left\vert {\mathcal J},\mu \right\rangle $  and  $%
\left\vert {\tilde {\mathcal J}},\mu \right\rangle $.  Therefore, below we will omit the tilde sign and use notations
$\vert {\mathcal J},\mu \rangle $  in a general sense.

In the  $\vert {\mathcal J},\mu \rangle $  basis,  the Hamiltonian matrix
is given by
\begin{eqnarray}\label{hamnottilde}
&&\langle {\mathcal J},\mu ^{\prime }|H_{int+\lambda,\Delta }|{\mathcal J},\mu\rangle
=\epsilon _{\mu }\delta _{\mu ^{\prime }\mu
}+\\\nonumber
&&\sum_{n=1}^{15}E_{n}\langle {\mathcal J},\mu ^{\prime }|{\mathcal I},n\rangle
\langle {\mathcal I},n|{\mathcal J},\mu \rangle
\end{eqnarray}
where $\epsilon _{\mu }$  is the energy of the $\vert {\mathcal J},\mu \rangle $ state,  and the $\langle {\mathcal J},\mu \vert{\mathcal I},n\rangle$ denote  components of the overlap matrix.
  The diagonalization of (\ref{hamnottilde}) gives energy eigenstates of the full Hamiltonian
\begin{eqnarray}
\vert D,\xi \rangle =\sum_{\mu =1}^{15}c_{\xi \mu
}\vert {\mathcal J},\mu \rangle~,
\end{eqnarray}
where $\xi =1,...15$ and  $c_{\xi \mu }$ denote the eigenvectors. We denote  the energy eigenvalues   as $E_{\xi }$.
The
 block structure of the Hamiltonian matrix is given in Appendix A. As a final remark, we also note  that in the limit $J_H=0$ and  $\Delta=0$, the Hamiltonian matrix (\ref{hamnottilde}) is diagonal with $E_1=...=E_8=-\lambda /2 +6 U_2$, $E_9=-2\lambda+6 U_2$, and $E_{10}=...=E_{15}=\lambda +6 U_2$.

\section{Derivation of the super-exchange Hamiltonian}\label{sec:ham}

In systems  with tetragonal symmetry, the Ir-O-Ir bonds  are close to 180$^\circ$. In these systems, in general,  the contribution to the super-exchange coupling from direct Ir-Ir hopping may be neglected  because the Ir ions are quite far from each other. The dominant contribution to the super-exchange is from the hopping via  intermediate oxygen ions, so-called oxygen-assisted hopping.   Because intermediate states with two holes on the oxygen ion have high-energy and, thus, can be neglected, we may integrate out
the  oxygen degrees of freedom to obtain an effective oxygen-assisted hopping between Ir 5$d$-states.
 Then applying a second order perturbation theory expansion in the effective hopping
parameters, we obtain a super-exchange
Hamiltonian  in the following form:
\begin{eqnarray}
H_{ex,n,n^{\prime }}=\sum_{\xi }\frac{1}{\epsilon _{\xi }}PH_{t,n,n^{\prime
}}Q_{\xi ,n^{\prime }}H_{t,n^{\prime },n}P~,
\end{eqnarray}
where
\begin{eqnarray}
P=\sum_{\sigma
_{n}=\pm 1}\vert 1/2,\sigma _{n}/2;n\rangle \langle
n;1/2,\sigma_{n} /2\vert
\end{eqnarray}
is the projection operator onto the  ground states with one hole at site $n$. The projection
operators onto two-hole intermediate states $\vert D,\xi;n^{\prime} \rangle $
with
excitation energy $\epsilon_{\xi }$ at site $n^{\prime}$ are then given by
 \begin{eqnarray}
 Q_{\xi ,n^{\prime}}=\vert D,\xi ;n^{\prime}\rangle \langle n^{\prime};D,\xi
\vert = D_{\xi ,n^{\prime}}^\dagger D_{\xi ,n^{\prime}}~.
\end{eqnarray}
 The excitation energies
of the intermediate states are  $\epsilon _{\xi }=E_{0h}+E_{\xi }-2E_{1h}$.

The connection between the Kramers doublet ground states  at site $n$ ($\gamma =1,2$) and the full manifold of states at site $n^{\prime}$ ($\gamma^{\prime} =1,2,...,6$)
is  given by the projected hopping term:
\begin{eqnarray}
PH_{t,n,n^{\prime}}=\sum_{\gamma =1}^{2}\sum_{\gamma ^{\prime
}=1}^{6}T_{n,n^{\prime}}^{\gamma ,\gamma ^{\prime }}b_{n,\gamma }^\dagger b_{n^{\prime},\gamma
^{\prime }}~,
\end{eqnarray}
where the elements of the matrix $T_{n,n^{\prime}}^{\gamma ,\gamma ^{\prime }}$
describe an overlap between   $\vert J, J_z\rangle$  or $\vert \Psi_{\gamma}\rangle$ states in the absence  or in the presence of  the tetragonal distortion, respectively.
Explicitly,  these matrices are derived in Appendix B.
Next, we apply $PH_{t,n,n^{\prime}}$ to the $D_{\xi ,n^{\prime}}^\dagger$ state and obtain
\begin{eqnarray}
&&P H_{t,n,n^{\prime}}D_{\xi ,n^{\prime}}^\dagger =\\\nonumber
&&\sum_{\gamma
=1}^{2}\sum_{\gamma ^{\prime }=1}^{6}T_{n,n^{\prime}}^{\gamma ,\gamma
^{\prime }}b_{n,\gamma }^\dagger b_{n^{\prime},\gamma ^{\prime }}
\sum_{\nu
=1}^{15}\sum_{\gamma _{1},\gamma _{2}=1}^{6}c_{\xi ,\nu }m_{\gamma
_{1}\gamma _{2}}^{\nu }b_{n^{\prime},\gamma _{1}}^\dagger b_{n^{\prime},\gamma _{2}}^\dagger
\\\nonumber
&&=\sum_{\gamma,\gamma ^{\prime }=1}^{2}\sum_{\gamma
_{1}=1}^{6} \sum_{\nu =1}^{15} T_{n,n^{\prime}}^{\gamma ,\gamma
_{1}}c_{\xi ,\nu }(m_{\gamma _{1}\gamma ^{\prime }}^{\nu }-m_{\gamma
^{\prime }\gamma _{1}}^{\nu })b_{n,\gamma }^\dagger b_{n^{\prime},\gamma ^{\prime }}^\dagger~.
\end{eqnarray}
Here the terms with $b_{n^{\prime},\gamma _{1}}^\dagger b_{n^{\prime},\gamma _{2}}^\dagger$ for
$\gamma _{1},\gamma _{2} > 2$ are projected out by the operator $P_{n^{\prime}}$.
Finally, using the  following relation:
\begin{eqnarray}
PH_{t}Q_{\xi ,n^{\prime}}H_{t}P=[PH_{t}D_{\xi ,n^{\prime}}^\dagger][PH_{t}D_{\xi
,n^{\prime}}^\dagger]^\dagger,
\end{eqnarray}
where
\begin{eqnarray}
PH_{t,n,n^{\prime}}D_{\xi ,n^{\prime }}^\dagger =\sum_{\sigma ,\sigma ^{\prime }=\pm
1}A_{\sigma ,\sigma ^{\prime }}^{\xi }b_{n,\sigma }^\dagger b_{n^{\prime},\sigma
^{\prime }}^\dagger ~
\end{eqnarray}
with
\begin{eqnarray}\label{def:A}
A_{n,n^{\prime};\sigma ,\sigma ^{\prime }}^{\xi }=\sum_{\gamma _{1}=1}^{6}
\sum_{\nu =1}^{15}T_{n,n^{\prime}}^{\sigma ,\gamma _{1}}c_{\xi ,\nu
}(m_{\gamma _{1}\sigma ^{\prime }}^{\nu }-m_{\sigma ^{\prime }\gamma
_{1}}^{\nu })~,
\end{eqnarray}
 we  write the exchange Hamiltonian as
\begin{eqnarray}\label{hamex}
H_{ex,n,n^{\prime}}&=&\sum_{\sigma ,\sigma ^{\prime }=1}^{2}\sum_{\sigma
_{1},\sigma _{1}^{\prime }=1}^{2}\sum_{\xi =1 }^{15}\\\nonumber
&&\frac{1}{\epsilon _{\xi }}%
\{A_{n,n^{\prime};\sigma ,\sigma ^{\prime }}^{\xi }b_{n,\sigma }^\dagger b_{n^{\prime},\sigma
^{\prime }}^\dagger A_{n^{\prime},n;\sigma _{1}^{\prime },\sigma _{1}}^{\xi
}b_{n^{\prime},\sigma _{1}^{\prime }}b_{n,\sigma _{1}}\}~.
\end{eqnarray}
We note that $ A_{n^{\prime},n;\sigma^{\prime} ,\sigma }^{\xi }=\left(A_{n,n^{\prime};\sigma ,\sigma ^{\prime }}^{\xi }\right)^*$. In the following, in order to shorten notations, we omit the site indices denoting
$A_{n,n^{\prime};\sigma ,\sigma ^{\prime }}^{\xi }\equiv A_{\sigma ,\sigma ^{\prime }}^{\xi }$ and $A_{n^{\prime},n;\sigma^{\prime} ,\sigma }^{\xi }\equiv \left(A_{\sigma ,\sigma ^{\prime }}^{\xi }\right)^*$.
 We also observe that $\sum_{\xi }\frac{1}{\epsilon _{\xi }}A_{\sigma ,\sigma }^{\xi }A_{\sigma_{1},-\sigma _{1}}^{\xi }=0$, since $A_{\sigma ,\sigma }^{\xi }$ and $A_{\sigma _{1},-\sigma _{1}}^{\xi }$ connect different groups of states $\left\vert D,\xi \right\rangle $ and are therefore "orthogonal".
Defining operators $B_{n\sigma \sigma ^{\prime }}=b_{n,\sigma
}^\dagger b_{n,\sigma ^{\prime }}$, we may write the superexchange Hamiltonian (\ref{hamex}) in the form
\begin{eqnarray}\label{hamex3}
&&H_{ex,n,n^{\prime}}=\sum_{\xi=1 }^{15}\frac{1}{\epsilon _{\xi }}\{\\\nonumber
 &&
A_{\uparrow\uparrow }^{\xi }\left(A_{\uparrow \uparrow }^{\xi }\right)^*\bigl(
B_{n\uparrow \uparrow}B_{n^{\prime}\uparrow \uparrow }+B_{n\downarrow \downarrow }B_{n^{\prime}
\downarrow \downarrow }\bigr)+\\\nonumber
&&A_{\uparrow \downarrow }^{\xi }\left(A_{\uparrow
\downarrow }^{\xi }\right)^*\bigl(
B_{n\uparrow \uparrow }B_{n^{\prime}\downarrow
\downarrow }+B_{n\downarrow \downarrow }B_{n^{\prime} \uparrow \uparrow }\bigr)+\\\nonumber
&&
A_{\uparrow
\downarrow }^{\xi }\left(A_{\downarrow \uparrow }^{\xi }\right)^*\bigl(
B_{n\uparrow \downarrow}B_{n^{\prime} \downarrow \uparrow }+B_{n\downarrow \uparrow }
B_{n^{\prime}\uparrow \downarrow }\bigr)+\\\nonumber
&& A_{\uparrow \uparrow }^{\xi }\left(A_{\downarrow
\downarrow }^{\xi }\right)^*\bigl(B_{n\uparrow \downarrow }B_{n^{\prime}\uparrow
\downarrow }+B_{n\downarrow \uparrow }B_{n^{\prime}\downarrow \uparrow }\bigr)\}
\end{eqnarray}

\begin{figure*}\label{anisoplots2}
\includegraphics[width=0.95\columnwidth]{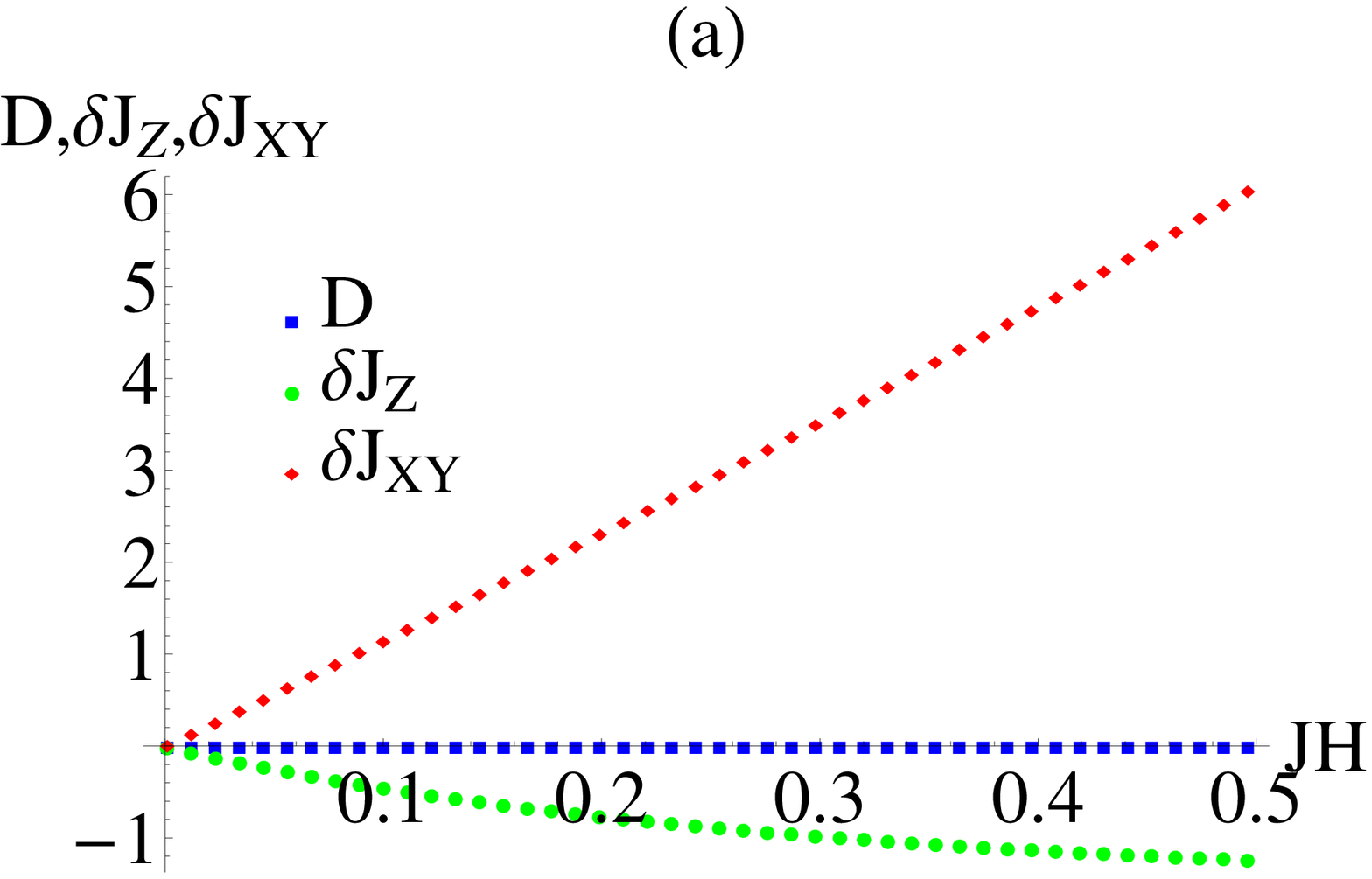}
\includegraphics[width=0.95\columnwidth]{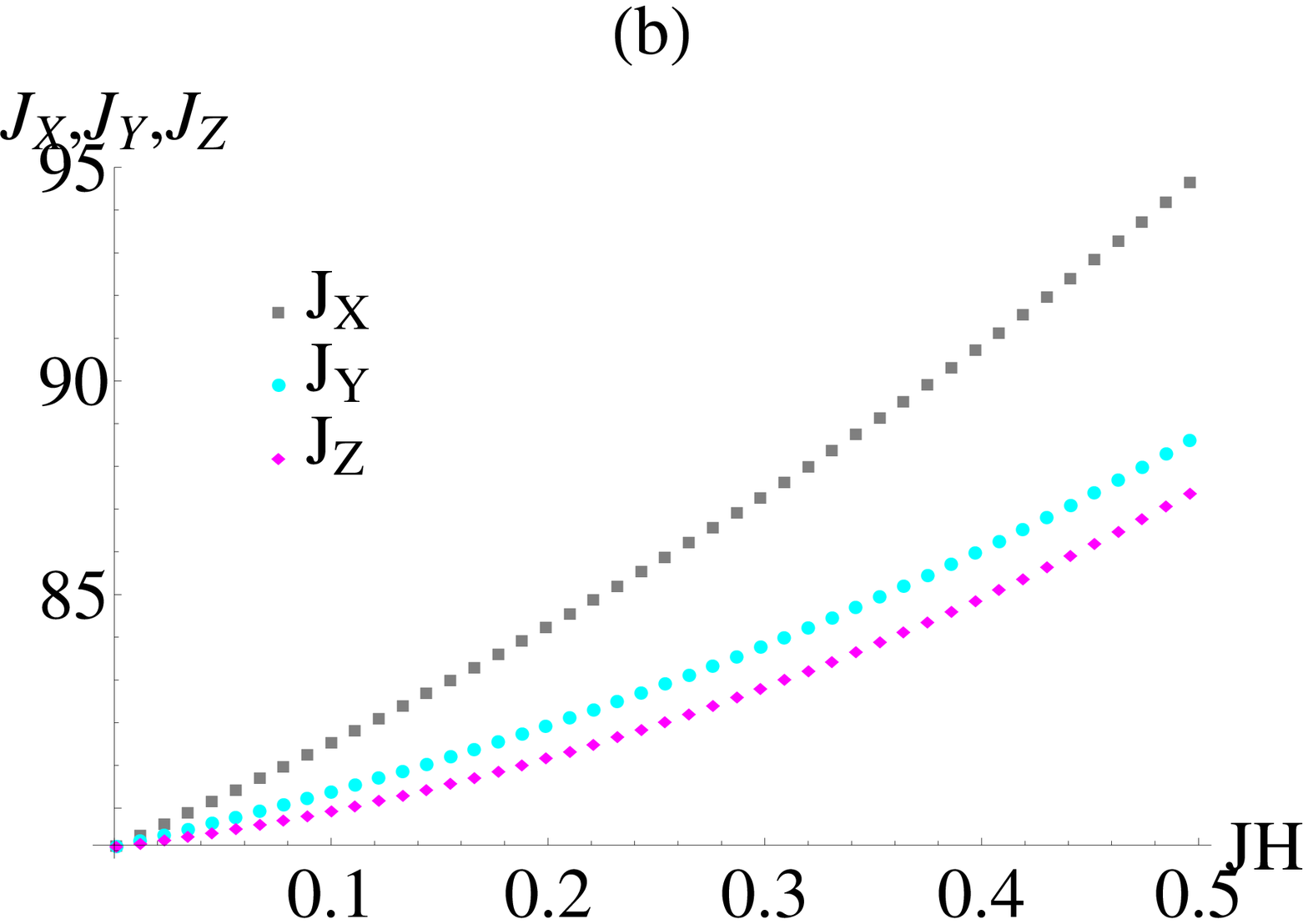}
\caption{(a) Anisotropic exchange couplings $\delta J_{xy},\delta J_{z}$ and the DM constant $D$ in meV (shown by red diamond, green circle and blue square lines, respectively) as functions of Hund's coupling, $J_H$. (b)
The  exchange couplings $J_{x},J_{y},J_{z}$ in meV (shown by  gray square, cyan circle and magenta diamond lines, respectively) as
functions of Hund's coupling, $J_H$.
 The  microscopic parameters of the model are considered to be $\alpha=0$ rad, $\Delta=0.15$ eV, $U_2=1.8$ eV,
 $\lambda=0.4$ eV, $t_{eff}=0.13$ eV. }
\end{figure*}
Next, we
introduce  pseudospin operators $S_{n}^{\alpha }=\frac{1}{2}\sum_{\sigma
,\sigma ^{\prime }=\pm 1}\tau _{\sigma ,\sigma ^{\prime }}^{\alpha
}b_{\sigma ,n}^\dagger b_{\sigma ^{\prime },n}$  with the Pauli matrices   $\tau _{\sigma ,\sigma ^{\prime }}^{\alpha}, \alpha =x,y,z$
and $\rho _{n}=\sum_{\sigma =\pm 1}b_{\sigma ,n}^\dagger b_{\sigma ,n}$ and express
operators $B_{n\sigma \sigma ^{\prime }}$  in terms of pseudospin operators as
\begin{eqnarray}
B_{n\uparrow \uparrow}&=&S_{n}^{z }+\rho _{n}\nonumber\\
B_{n\downarrow \downarrow}&=&-S_{n}^{z }+\rho _{n}\nonumber\\
B_{n\uparrow \downarrow}&=&S_{n}^{+ }\\\nonumber
B_{n\downarrow \uparrow}&=&S_{n}^{-}.
\end{eqnarray}
 This  allows us to write the super-exchange Hamiltonian  (\ref{hamex3}) on the bond $n,n^{\prime}$  in  terms
of the magnetic  degrees of freedom of Ir$^{4+}$ as
\begin{eqnarray}\label{spinham}
H_{ex,n,n^{\prime}}=&&J_{z}S_{n}^{z}S_{n^{\prime}}^{z}+J_{x}S_{n}^{x}S_{n^{\prime}}^{x}+
J_{y}S_{n}^{y}S_{n^{\prime}}^{y}\\\nonumber
&&-D(S_{n}^{x}S_{n^{\prime}}^{y}-S_{n}^{y}S_{n^{\prime}}^{x})+W \rho
_{n}\rho _{n^{\prime}}~,
\end{eqnarray}
where the coupling constants are given by the following expressions:
\begin{eqnarray}
 J_{z}=-2\sum_{\xi }\frac{1}{\epsilon _{\xi }}&&\Bigl( A_{\uparrow \uparrow
}^{\xi }\left(A_{\uparrow \uparrow }^{\xi }\right)^*+ A_{\downarrow \downarrow
}^{\xi }\left(A_{\downarrow \downarrow }^{\xi }\right)^*\\\nonumber
&&- A_{\uparrow \downarrow
}^{\xi }\left(A_{\uparrow \downarrow }^{\xi }\right)^*- A_{\downarrow \uparrow
}^{\xi }\left(A_{\downarrow \uparrow }^{\xi }\right)^*\Bigr),
\end{eqnarray}
\begin{eqnarray}
J_{x}=-2\sum_{\xi }\frac{1}{\epsilon _{\xi }}&&\Bigl( A_{\uparrow \uparrow
}^{\xi }\left(A_{\downarrow \downarrow }^{\xi }\right)^*+ A_{\downarrow \downarrow
}^{\xi }\left(A_{\uparrow \uparrow }^{\xi }\right)^*\\\nonumber &&+ A_{\uparrow \downarrow
}^{\xi }\left(A_{\downarrow \uparrow }^{\xi }\right)^*+ A_{\downarrow \uparrow
}^{\xi }\left(A_{\uparrow \downarrow }^{\xi }\right)^*\Bigr),
\end{eqnarray}
\begin{eqnarray}
J_{y}=2\sum_{\xi }\frac{1}{\epsilon _{\xi }}&&\Bigl( A_{\uparrow \uparrow
}^{\xi }\left(A_{\downarrow \downarrow }^{\xi }\right)^*+ A_{\downarrow \downarrow
}^{\xi }\left(A_{\uparrow \uparrow }^{\xi }\right)^*\\\nonumber&&- A_{\uparrow \downarrow
}^{\xi }\left(A_{\downarrow \uparrow }^{\xi }\right)^*- A_{\downarrow \uparrow
}^{\xi }\left(A_{\uparrow \downarrow }^{\xi }\right)^*\Bigr),
\end{eqnarray}
\begin{eqnarray}
D=2\imath \sum_{\xi }\frac{1}{\epsilon _{\xi }}\Bigl( A_{\uparrow \downarrow
}^{\xi }\left(A_{\downarrow \uparrow }^{\xi }\right)^*- A_{\downarrow \uparrow
}^{\xi }\left(A_{\uparrow \downarrow }^{\xi }\right)^*\Bigr),
\end{eqnarray}
\begin{eqnarray}
W=-2\sum_{\xi }\frac{1}{\epsilon _{\xi }}&&\Bigl( A_{\uparrow \uparrow
}^{\xi }\left(A_{\uparrow \uparrow }^{\xi }\right)^*+ A_{\downarrow \downarrow
}^{\xi }\left(A_{\downarrow \downarrow }^{\xi }\right)^*+\nonumber\\&&
 A_{\uparrow \downarrow
}^{\xi }\left(A_{\uparrow \downarrow }^{\xi }\right)^*+ A_{\downarrow \uparrow
}^{\xi }\left(A_{\downarrow \uparrow }^{\xi }\right)^*\Bigr)
\end{eqnarray}
The last interaction term W gives a constant energy shift and we will omit it. It is also convenient to rewrite the remaining terms introducing the following notations: $\delta J_z=J_z-J_y$, $\delta J_{xy}=J_x-J_y$ on  $x$-bond and
$\delta J_z=J_z-J_x$, $\delta J_{xy}=J_y-J_x$ on $y$-bond.  Then, $H_{ex,n,n^{\prime}}$  can be written as
\begin{eqnarray}
\label{hamtet}
&&H_{ex,n,n^{\prime}}=J {\bf S}_n {\bf S}_{n^{\prime}}-D(S_{n}^{x}S_{n^{\prime}}^{y}-S_{n}^{y}S_{n^{\prime}}^{x})
\\\nonumber
&&+\delta J_z S^z_n  S^z_{n^{\prime}}
+\delta J_{xy}\left({\bf S}_n \cdot {\bf r}_{n,n^{\prime}}\right)
 \left({\bf S}_{n^{\prime}} \cdot{\bf r}_{n,n^{\prime}}\right)~,
\end{eqnarray}
where ${\bf r}_{n,n^{\prime}}$ is the unit vector along the $n,n^{\prime}$ bond.
In this form the nature of interactions between pseudospin moments $S$ is more clear.  The first term describes the Heisenberg isotropic interaction with a coupling constant $J=J_y$ on the $x$-bond.  We note that for any possible set of microscopic parameters, the isotropic exchange is the dominant exchange  and has AFM nature. The second term is a Dzyaloshinsky-Moriya (DM) interaction with a coupling constant  $D$, which leads to a spin canting in the $xy$-plane proportional to the ratio $D/J$.
 The third term
describes an additional Ising-like interaction of $z$-components of spins.  $\delta J_z >0$  favors AFM ordering of spins along the $z$ axis and works as an easy axis anisotropy. $\delta J_z <0$ supports  FM ordering of spins along the $z$ axis and works as an easy plane anisotropy. The last term is a pseudo-dipolar interaction.
 Finally, the total superexchange  Hamiltonian  is given by
\begin{eqnarray}
H=\sum_{\langle n,n^{\prime}\rangle}H_{ex,n,n^{\prime}},
\end{eqnarray}
where summation is over all bonds of the lattice.

\begin{figure*}\label{anisoplots3}
\includegraphics[width=0.95\columnwidth]{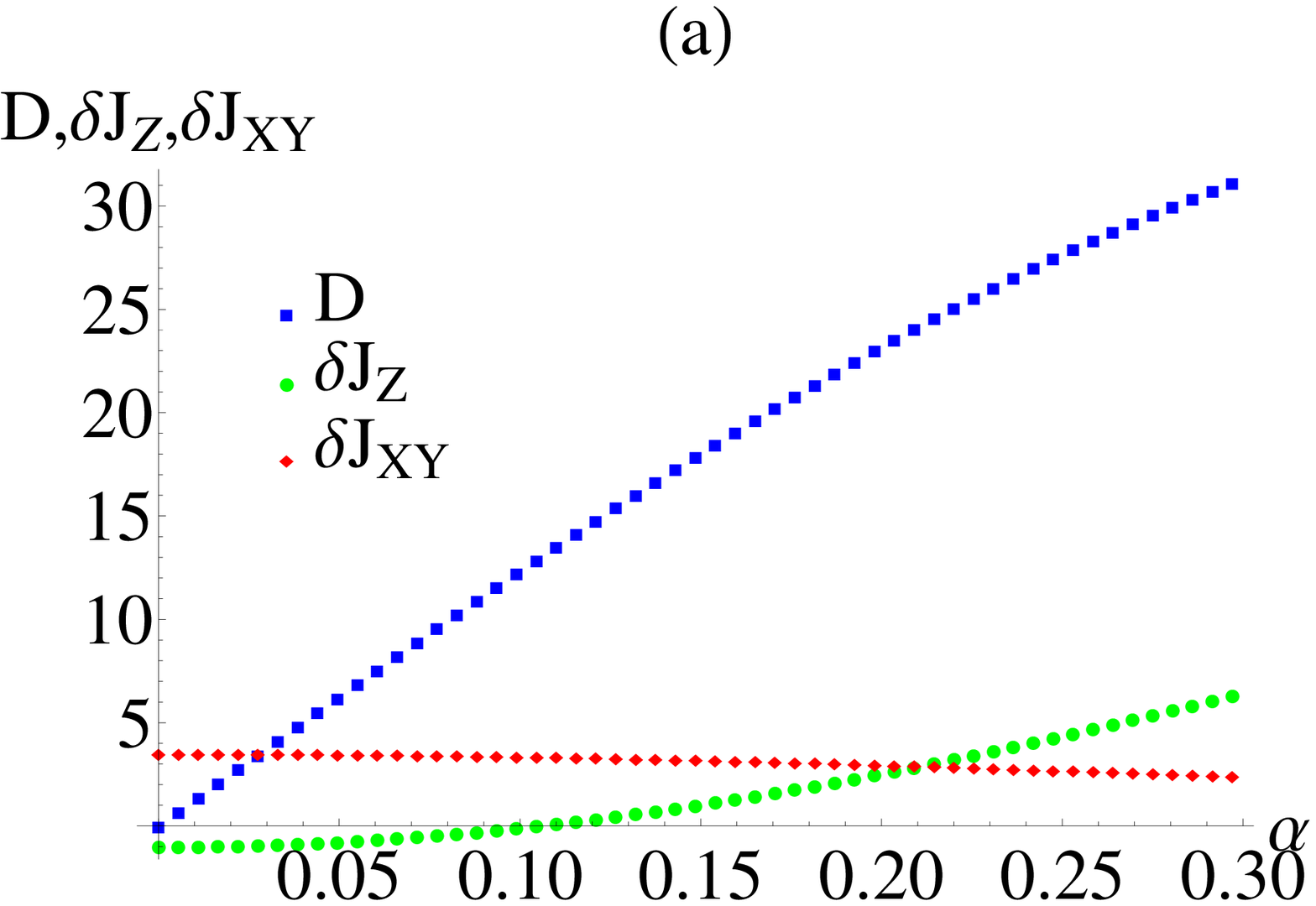}
\includegraphics[width=0.95\columnwidth]{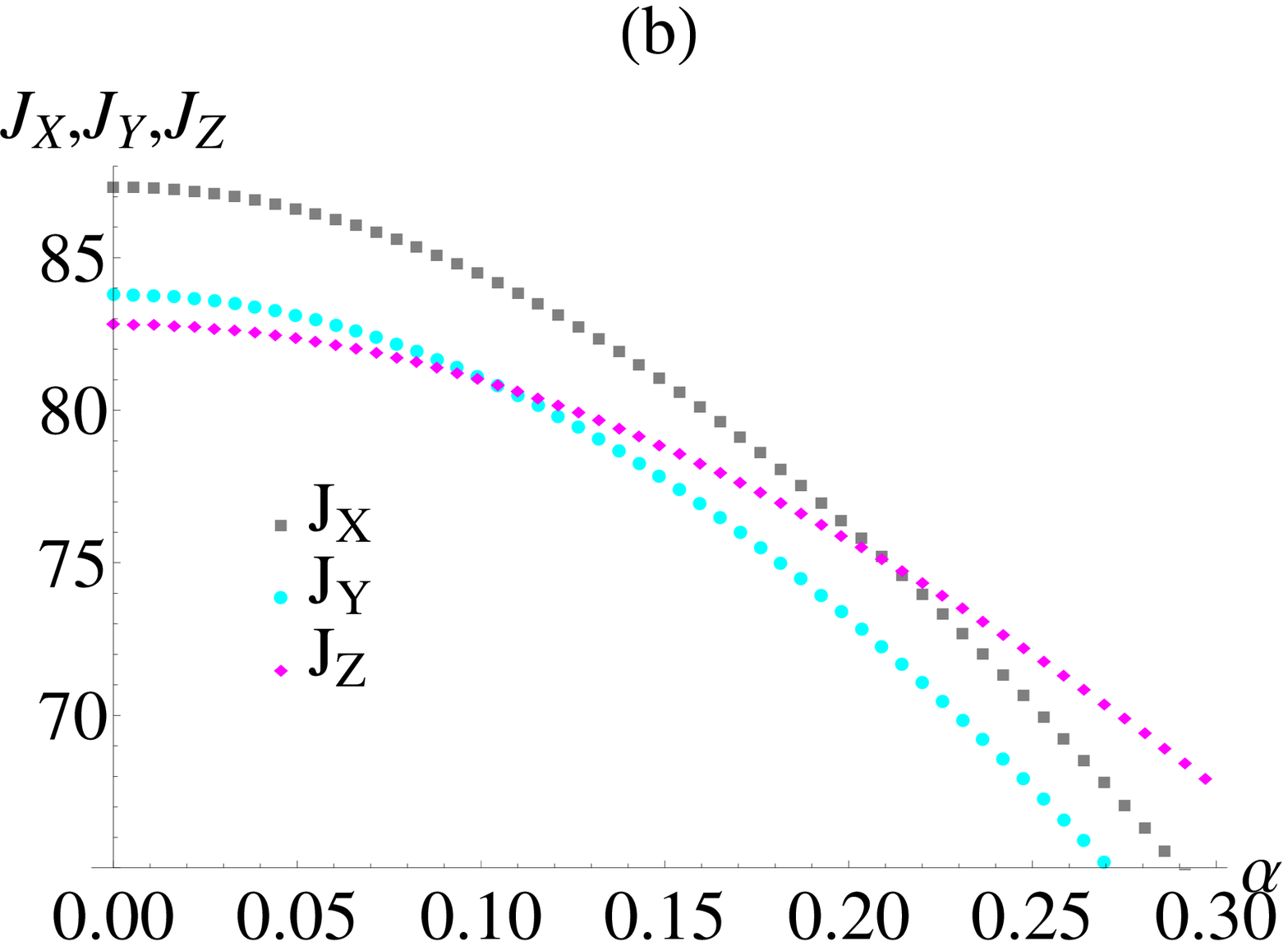}
\caption{
 (a) Anisotropic exchange couplings $\delta J_{xy},\delta J_{z}$ and DM constant $D$  in meV (shown by red diamond, green circle and blue square lines, respectively)    and  (b) the  exchange couplings $J_{x},J_{y},J_{z}$ in meV (shown by  gray square, cyan circle and magenta diamond lines, respectively) as
functions of the rotation angle $\alpha$. The other microscopic parameters considered to be $\Delta=0.15$ eV, $U_2=1.8$ eV, $J_H=0.3$ eV,
 $\lambda=0.4$ eV and  $t_{eff}=0.13$ eV. }
\end{figure*}

\section{Results and discussions}\label{sec:results}

\subsection{ Application to Sr$_2$IrO$_4$}

Below we present our results on how the exchange coupling constants $J_x,J_y, J_z$  and anisotropic couplings $\delta J_{xy},\delta J_z$ and $D$ depend on the microscopic parameters of the system.
We first note that  the main role of the Coulomb repulsion is to determine the overall energy scale for  the couplings.  In all computations we take $U_2=1.8$ eV, which lays inside the range of values, 1.5 eV-2.5 eV,  characteristic  for iridates.\cite{martins11,arita12}  We  will mostly set the SOC constant to be equal to $\lambda=0.4$ eV - the value  associated with Ir ions in the literature, however, we  will also consider the smaller value   $\lambda=0.22$ eV, which  was suggested in the experimental work by Haskel {\it et al}.~\cite{haskel12}

    Sr$_2$IrO$_4$ is also characterized  by various structural distortions,  the most important of which are  the tetragonal distortion and rotations of the oxygen octahedra. Both of them are present  even at ambient pressure. In calculations we  either consider the tetragonal distortion  and the angle of rotation to be equal to $\Delta=0.15$ eV and $\alpha=0.2$ rad, respectively, or we study how the exchange parameters depend on these quantities.

 Finally, we consider the hopping parameter  between Ir ions to be equal to $t_{eff}=0.13$ eV, which is slightly lower than  the values 0.2-0.3 eV suggested by ab-initio calculations. These values of hoppings give  too large values of exchange couplings if all other parameters are set as we described above. We believe,  however, that hopping parameters obtained within density functional theory  are often  reduced when correlations are taken into account.

{\it The effect of Hund's coupling.} In Fig.2 (a) we plot the anisotropic couplings $\delta J_{xy},\delta J_z$ and the DM interaction constant $D$  as  functions of Hund's coupling, $J_{H}$, in the absence of  octahedral rotations, $\alpha=0$. In this case, the components of the vectors $A_{\sigma ,\sigma ^{\prime }}^{\xi }$ (\ref{def:A}) satisfy the following  conditions:
\begin{eqnarray}\label{condition}
\sum_{\xi }\frac{1}{\epsilon _{\xi }}(A_{\uparrow \uparrow }^{\xi
})^{2}&=&\sum_{\xi }\frac{1}{\epsilon _{\xi }}(A_{\downarrow \downarrow }^{\xi
})^{2}~,\nonumber\\
\sum_{\xi }\frac{1}{\epsilon _{\xi }}(A_{\uparrow \downarrow
}^{\xi })^{2}&=&\sum_{\xi }\frac{1}{\epsilon _{\xi }}(A_{\downarrow \uparrow
}^{\xi })^{2}~.
\end{eqnarray}
This symmetry reflects the fact that in the absence of the octahedra rotations there is no spin dependent hopping and,
 consequently, no DM interaction. The anisotropic terms $\delta J_z$ and $\delta J_{xy}$ are also zero at $\alpha=0$ and $J_H=0$ eV, but they acquire  finite values at $J_H\neq 0$. We note that the Ising-like interaction, $\delta J_z<0$,  makes  the $xy$-plane the pseudospin's easy plane.

 In Fig. 2 (b) we  plot  the exchange couplings $J_x,J_y,J_z$ as  functions of the Hund's coupling. On the $x$-bond,
 the isotropic exchange $J=J_y$. It is antiferromagnetic for all considered values of $J_H$ and its
 strength   varies in the range $J\in (78-95)$ meV for $J_H\in  (0-0.5)$ eV.
 This  compares well  not only with an estimate $J=51$ meV obtained by ab initio many-body quantum-chemical calculations\cite{{katukuri12}},
   but also with experimental findings in Sr$_2$IrO$_4$, for which  resonant inelastic x-ray scattering~\cite{kim12} and resonant magnetic
x-ray diffuse scattering measurements~\cite{fujiyama12}  indicate the isotropic exchange to be $J\simeq 60$ meV and $J\simeq 100$ meV, respectively.

\begin{figure*}\label{anisoplots4}
\includegraphics[width=0.98\columnwidth]{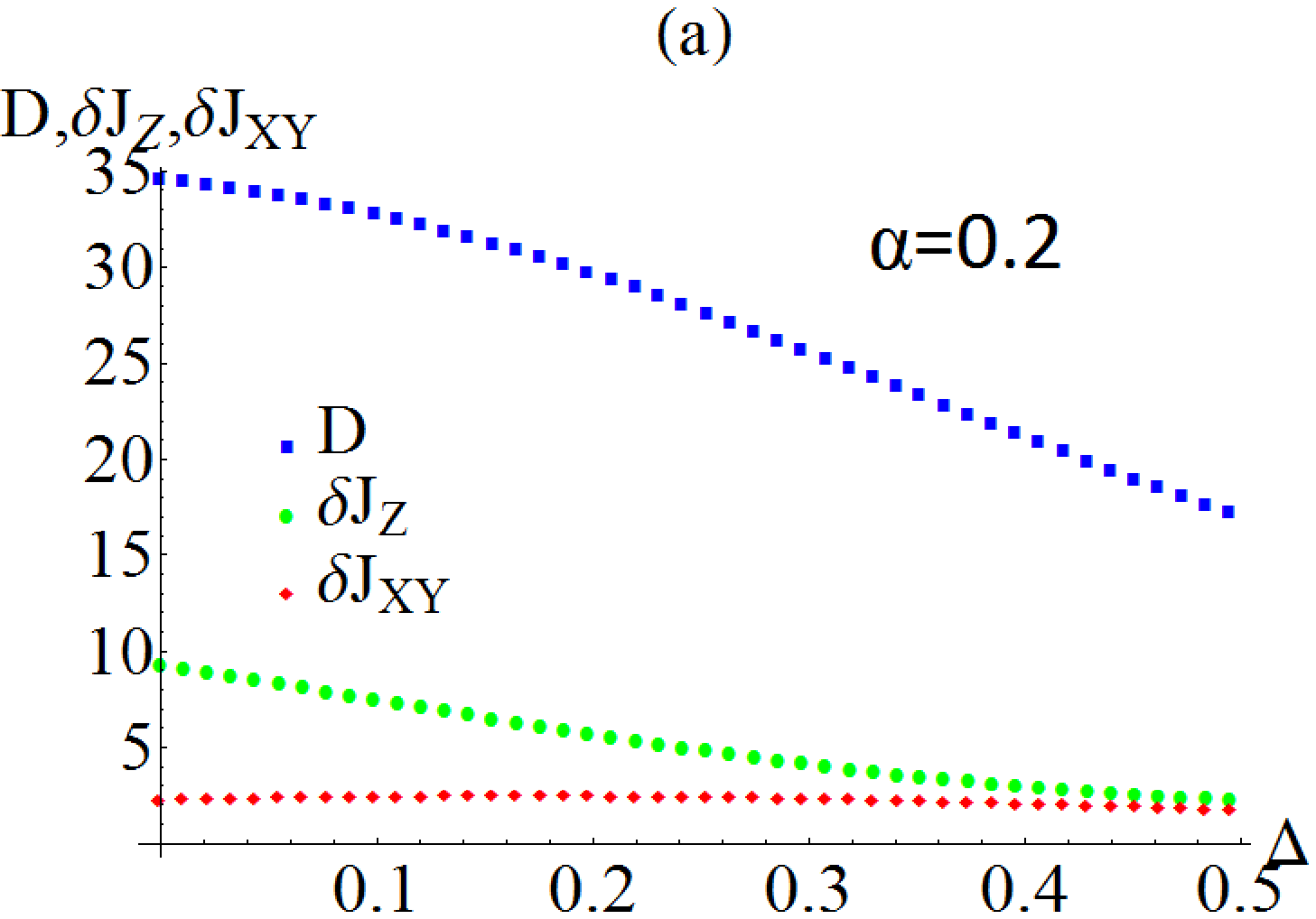}
\includegraphics[width=0.98\columnwidth]{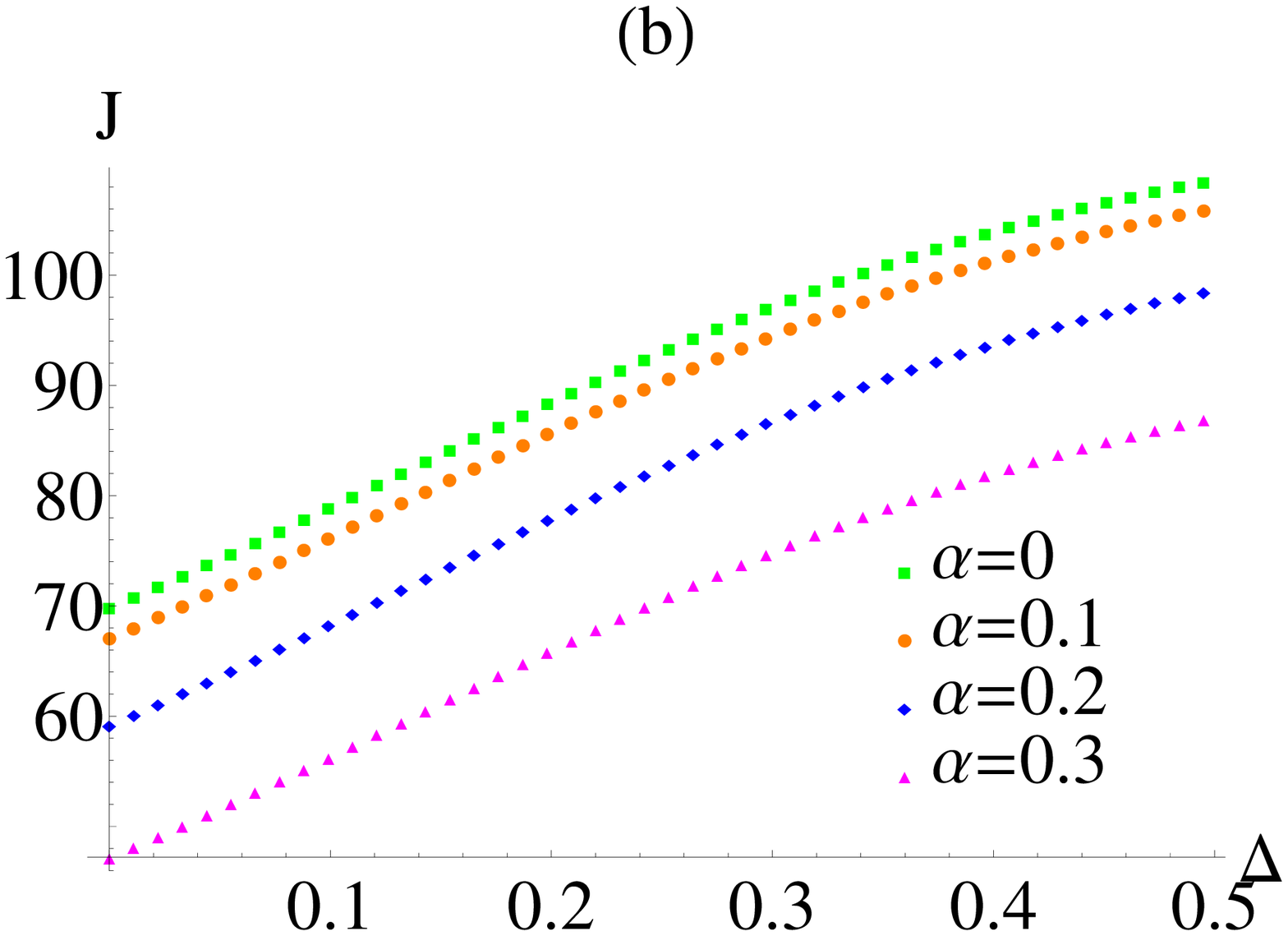}
\caption{(a) Anisotropic exchange couplings $\delta J_{xy},\delta J_{z}$ and DM constant $D$  in meV (shown by red diamond, green circle and blue square lines, respectively) as  functions of the tetragonal CF splitting computed for $\alpha=0.2$ rad
 (b) The dependencies of the isotropic exchange $J$ on the strength of the tetragonal CF splitting $\Delta$.
Green square, orange circle, blue diamond, magenta triangle lines correspond to $\alpha=0;0.1;0.2;0.3$ rad, respectively. The other parameters are $\lambda=0.4$ eV, $U_2=1.8$ eV, $J_H=0.3$ eV  and $t_{eff}=0.13$ eV.}
\end{figure*}

{\it The effect of staggered rotations of IrO$_6$ octahedra.}
The dependencies of the anisotropic couplings  $\delta J_{xy},\delta J_z$ and $D$  and the exchange constants  $J_x,J_y, J_z$  on the strength of the staggered rotations of the IrO$_6$ octahedra, $\alpha$, are presented in Fig.3 (a) and (b).  We first note that the isotropic exchange coupling $J=J_y\in (64-87)$ meV remains  in  good agreement with experimental estimates in the whole range of  values of $\alpha$ considered. However,  most  importantly,  the DM interaction becomes the dominant anisotropy
even at  small $\alpha$.
 At  $\alpha\simeq 0.2$ rad ($11.5^{\circ}$), the DM interaction  is already about $23$ meV, which roughly corresponds to third of the  isotropic interaction (see  Fig.3 (b)). Such  a large ratio between the DM interaction and the isotropic Heisenberg exchange  is very unusual and has never been observed in 3d transition metal oxides.

 The  other anisotropic interactions, both the pseudo-dipolar in-plane interaction, $\delta J_{xy}$, and  the Ising-like term, $\delta J_z$, remain relatively  small at finite values of $\alpha$.
  We note that $\delta J_z$ changes sign above some angle of octahedra rotation, $\alpha_c\simeq 0.1$ rad, but,  as it remains a subdominant interaction,  the magnetic moments remain lying in the
  $xy$-plane.

{\it The effect of tetragonal distortion.} Significant  changes in the super-exchange parameters  are caused by the tetragonal distortion. At ambient pressure the tetragonal distortion is about $\Delta\simeq 0.1$ eV, however, larger values can be easily reached under pressure.~\cite{haskel12} In Fig.4 (a) we plot  the dependencies of the anisotropic exchange couplings $\delta J_{xy},\delta J_{z}$ and  the DM coupling, $D$, on the strength of the tetragonal distortion, $\Delta$.
  An  increased tetragonal distortion leads to  a substantial decrease of both $D$ and $\delta J_{z}$,  but the overall hierarchy  of anisotropic interactions remains the same.  We see that if $\alpha$ was not changing under pressure, the  magnetic  anisotropy  would remain an easy plane anisotropy for all values of the tetragonal distortion.

  In Fig. 4 (b) we present  the results on how the isotropic exchange depends on  $\Delta$ at different values of  $\alpha$.
  We  see that the isotropic part of  the exchange coupling increases with increasing strength of the tetragonal distortion. We also  see that its dependence on $\Delta$  has a  quantitatively similar character for all $\alpha$,  with the largest  values  of $J$ reached at $\alpha=0$.  Importantly, for all values of $\alpha$ and $\Delta$, the isotropic exchange remains the dominant interaction with respect to the anisotropic terms.

\begin{figure*}\label{anisoplots5}
\includegraphics[width=0.68\columnwidth]{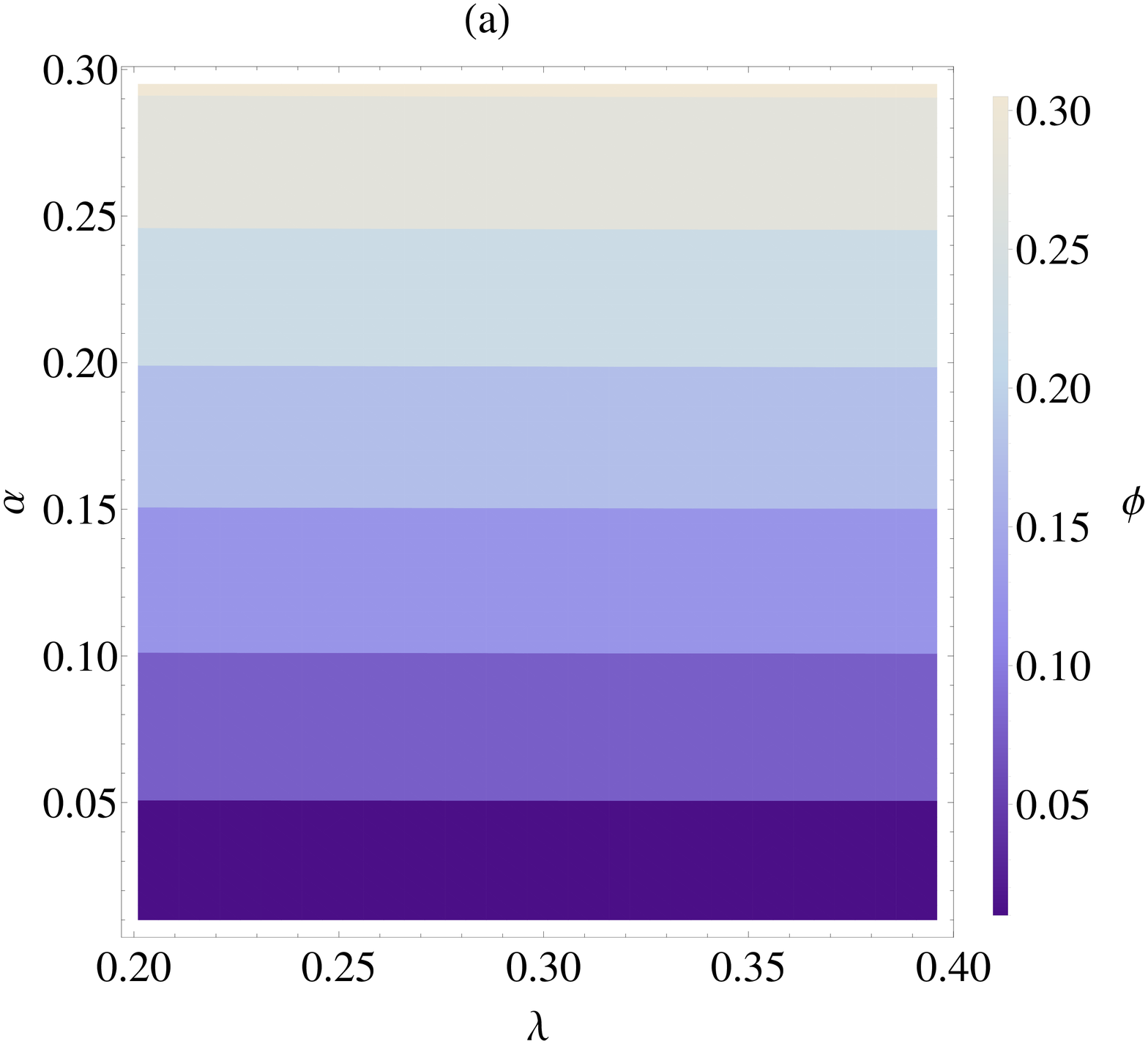}
\includegraphics[width=0.68\columnwidth]{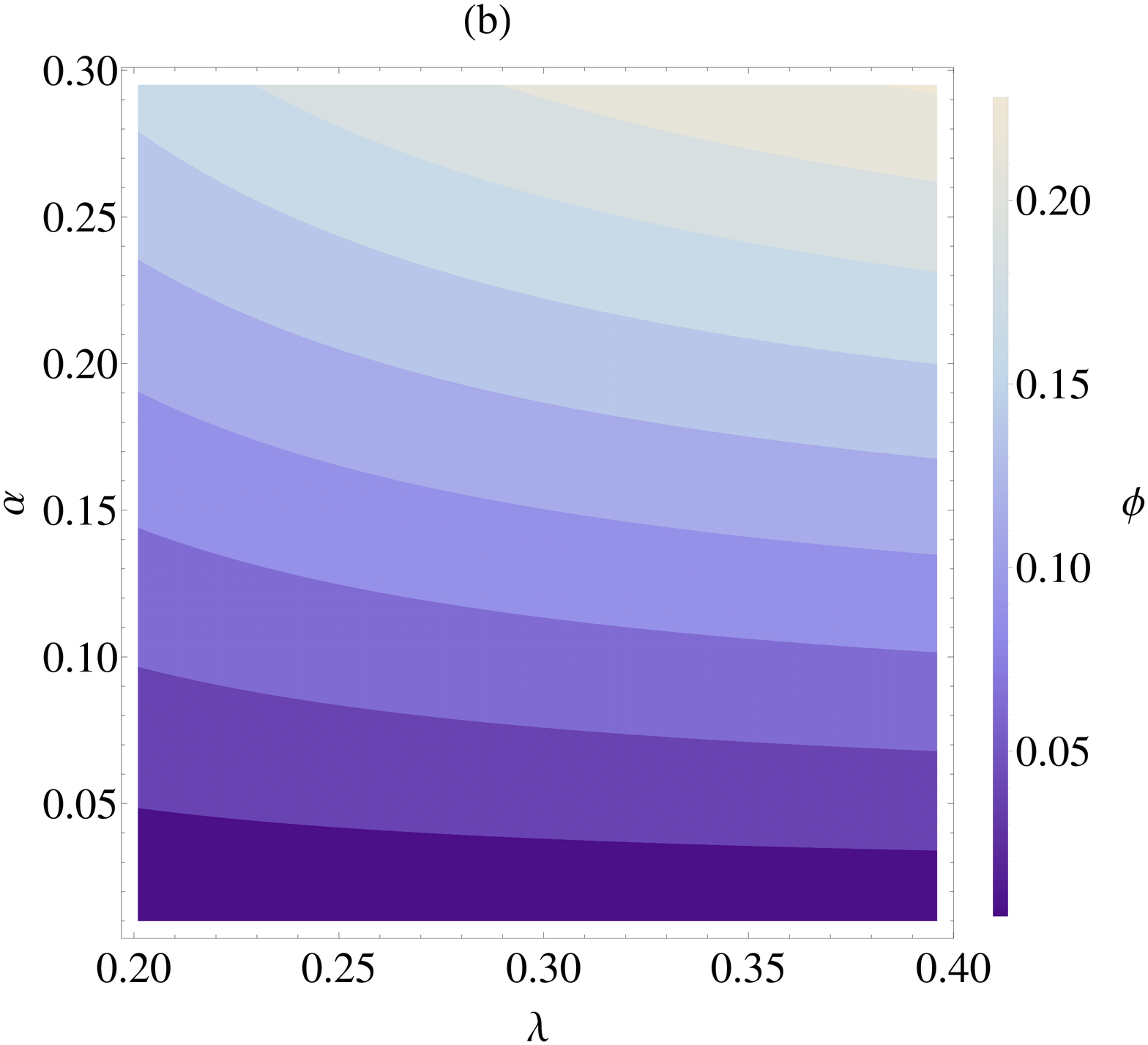}
\includegraphics[width=0.68\columnwidth]{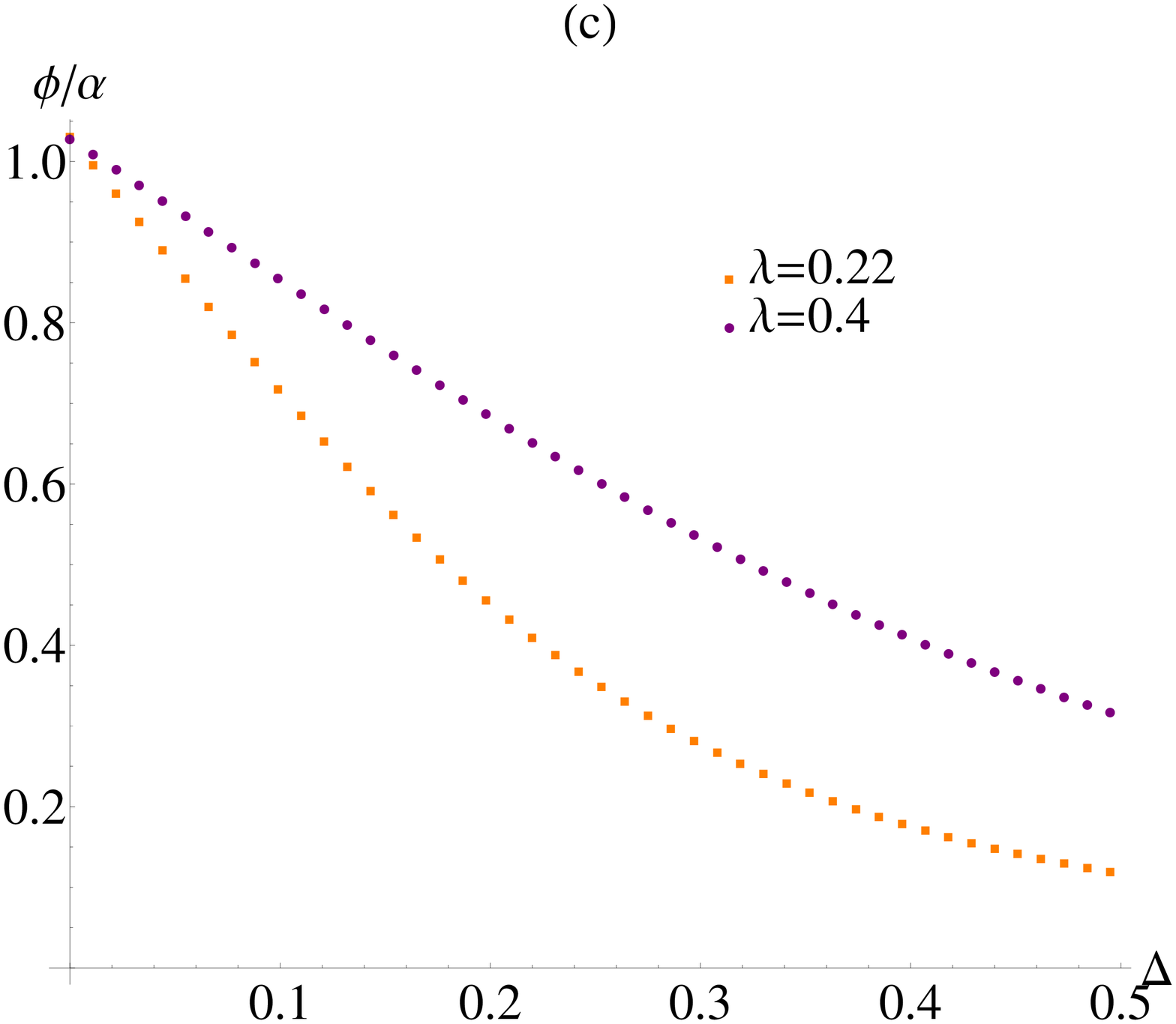}
\caption{
Mean field  magnetic phase diagram  in the parameter space of rotation angle $\alpha$ and the SOC  coupling $\lambda$ computed
(a) in the absence of the tetragonal distortion, $\Delta=0$ eV  and (b) in the presence of the tetragonal distortion, $\Delta=0.15$ eV.  In both parameter sets,  the obtained magnetic structure is  coplanar antiferromagnet with varying spin canting angle $\phi=[\pi-(\phi_A-\phi_B)]/2$, where $\phi_A$ and $\phi_B$ are the polar angles of spins on sublattices $A$ and $B$.  The color on the plots indicates the scale for which the angle $\phi$ changes with dark blue being the smallest and gray being the highest  value of the angle $\phi$.
The canted spin order is stabilized by a staggered rotations of oxygen tetrahedra in the presence of the SOC.  (c) The dependence of the  spin canting angle $\phi=1/2 \tan^{-1} \left( D/J \right)$	 (in units of $\alpha$) on the strength of the tetragonal CF splitting $\Delta$ computed for $\lambda=0.22$ eV (orange square line) and $\lambda=0.4$ eV (purple circle line).
The other microscopic parameters are chosen to be $U_2=1.8$ eV, $J_H=0.3$ eV, and  $t_{eff}=0.13$ eV, $\alpha=0.2$ rad. }
\end{figure*}

{\it Magnetic phase diagram.}
 Finally, we compute
a magnetic phase diagram of the  model (33).  The model allows for
two distinct magnetic phases: a coplanar  (or  collinear) two-sublattice antiferromagnet with spins lying in the $xy$-plane  and a collinear phase with spins pointing along the $c$-axis.
The coplanar phase is characterized by a  spin canting  angle $\phi$, which is  simply given by $\phi=1/2 \tan^{-1} \left( D/J \right)$. The dependence of the  spin canting angle $\phi$ (in units of $\alpha$)  on  $\Delta$  computed for $\alpha=0.2$  rad, corresponding to the angle of octahedral rotations at ambient pressure, and $J_H=0.3$ is presented  in Fig. 5(c).
We can see that in the cubic case, $\Delta=0$, the ratio $\phi/\alpha$  is equal to unity and, therefore, spins are canted exactly like the IrO$_6$ octahedra. At finite $\Delta$, the ratio $\phi/\alpha $ is smaller than one,  suggesting that in the presence of the tetragonal distortion  the spin structure  has an additional rigidity with respect to canting.

 A magnetic structure can be  determined by minimizing the classical energy taking into account  {\it all} exchange couplings present in the model. Assuming that
 in the presence of a staggered rotation of oxygen octahedra, the magnetic structure  is defined by two magnetic sublattices, A and B,  and that the orientation of the magnetic moments can be described with the help of four angles, $\theta_A,\theta_B,\phi_A,\phi_B$, we can write the classical energy as
 \begin{eqnarray}\label{classical1}
&&E_{cl} (\theta_A,\theta_B,\phi_A,\phi_B)
= J_z\cos\theta_A\cos\theta_B
  +\\\nonumber&&\frac{(J_x+J_y)}{2}\sin\theta_A\sin\theta_B(\cos\phi_A\cos\phi_B+
\sin\phi_A\sin\phi_B)
\\\nonumber&&-D\sin\theta_A\sin\theta_B(\cos\phi_A\sin\phi_B-\sin\phi_A\cos\phi_B)~.
\end{eqnarray}
 One can easily check that the contribution of the pseudo-dipolar interaction  to the classical energy cancels out. Thus, for any set of microscopic parameters, the classical ground state is determined by the competition between the DM interaction and the Ising-like anisotropy in the presence of a dominating AFM isotropic exchange.

In Fig.5 (a) and (b) we present a  mean field  magnetic phase diagram, where for each set of parameters the magnetic structure  is determined by minimizing $E_{cl}$  with respect to $\theta_A,\theta_B,\phi_A,\phi_B$.
 We considered two cases: Fig.5 (a) displays a phase diagram computed
 in the absence of tetragonal distortion ($\Delta=0$ eV)  and Fig.5 (b) displays a phase diagram computed in the presence of the tetragonal distortion ($\Delta=0.15$ eV). In both cases we considered the Hund's coupling to be equal to $J_H=0.3$ eV.  Both phase diagrams contain only the coplanar  antiferromagnet with varying canting angle $\phi=[\pi-(\phi_A-\phi_B)]/2$, where $\phi_A$ and $\phi_B$ are polar angles of spins on sublattices A and B.  The color on the plots indicates the magnitude scale of the angle $\phi$:  dark blue colors  correspond to the smallest and light gray colors correspond to the highest  values of the angle $\phi$.  As we discussed above, in the absence of the tetragonal  distortion, the canting of magnetic
 moments  rigidly follows the octahedral rotation and the canting angle $\phi$
 is exactly equal to the rotation angle $\alpha$  for all values of the  SOC strength (see Fig.5(a)). However, once the tetragonal distortion is present,  the canting angle $\phi$ becomes  smaller than $\alpha$. Moreover, the ratio $\phi/\alpha$ decreases with decrease of the SOC constant (see Fig.5(b)).

 We have to note that our findings are not in full agreement with the phase diagram presented by Jackeli and Khalliulin in Ref.\cite{jackeli09}, which shows that at large values of tetragonal distortion the spin-flop transition from
the in-plane canted spin state   happens to a collinear antiferromagnetic order  along the $z$-axis. Instead, we found that  at the considered set of parameters the tetragonal distortion may lead to a  disappearance of the ferromagnetic moment and a stabilization of the antiferromagnetic  order in the easy $xy$ plane.

 Our findings  are,  however, in qualitative agreement with both  pressure experiments  in Sr$_2$IrO$_4$~\cite{haskel12} and
  the x-ray resonant magnetic scattering study comparing  the magnetic and electronic structures of Sr$_2$IrO$_4$ and Ba$_2$IrO$_4$.~\cite{boseggia13}   The first study shows that when the tetragonal distortion due to pressure becomes relatively strong, about 17 GPa,
the  ferromagnetic order disappears. This magnetic transition was not  attributed to the gradual disappearance of the IrO$_6$ rotations under  pressure, because it would have likely resulted in  some kind of structural transition which was not observed. They also found that the application of  pressure up to at least 40 GPa does neither destroy the insulating behavior nor, probably,  the antiferromagnetic order. However, the direction of the  antiferromagnetic order parameter was not determined.
The second study shows the general robustness of the basal-plane antiferromagnetic order  in single-layer iridates. They found that
 in both Sr$_2$IrO$_4$ and Ba$_2$IrO$_4$, the antiferromagnetic component is oriented along the $[110]$ direction
 despite the fact that moving from Sr$_2$IrO$_4$ to Ba$_2$IrO$_4$ the tetragonal distortion is nearly doubled.

\begin{figure*}\label{anisoplots6}
\includegraphics[width=0.90\columnwidth]{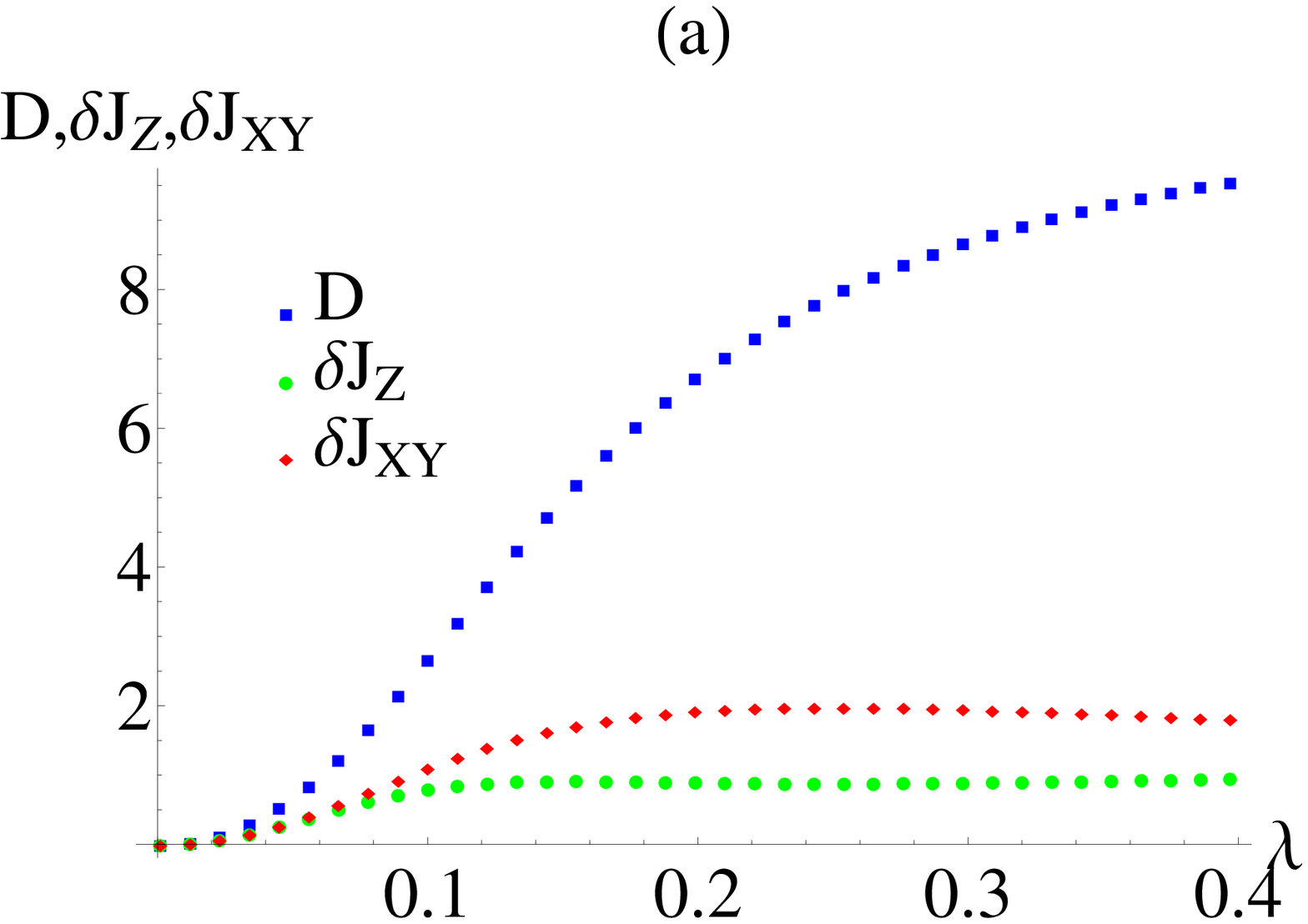}
\includegraphics[width=0.90\columnwidth]{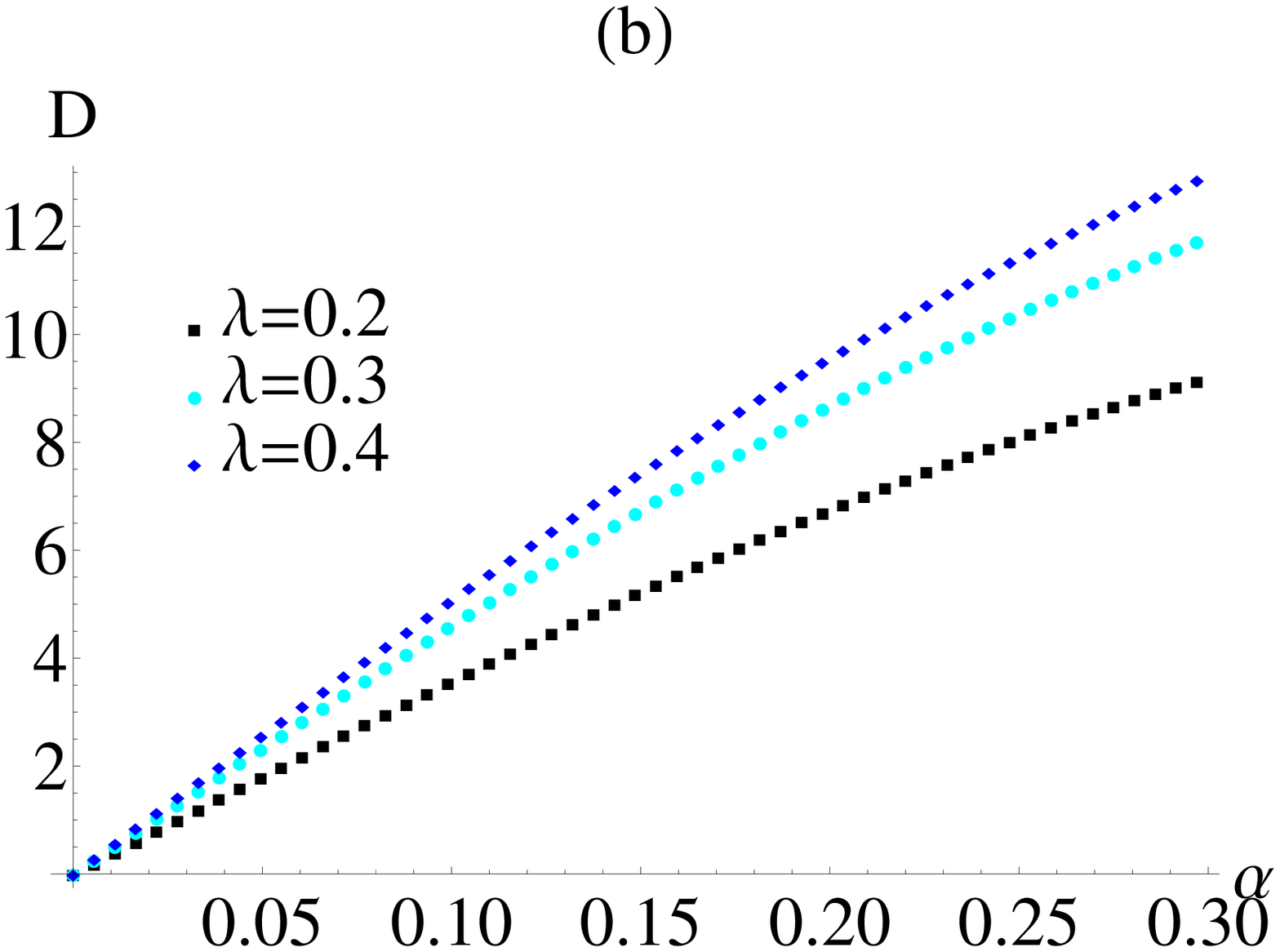}
\includegraphics[width=0.88\columnwidth]{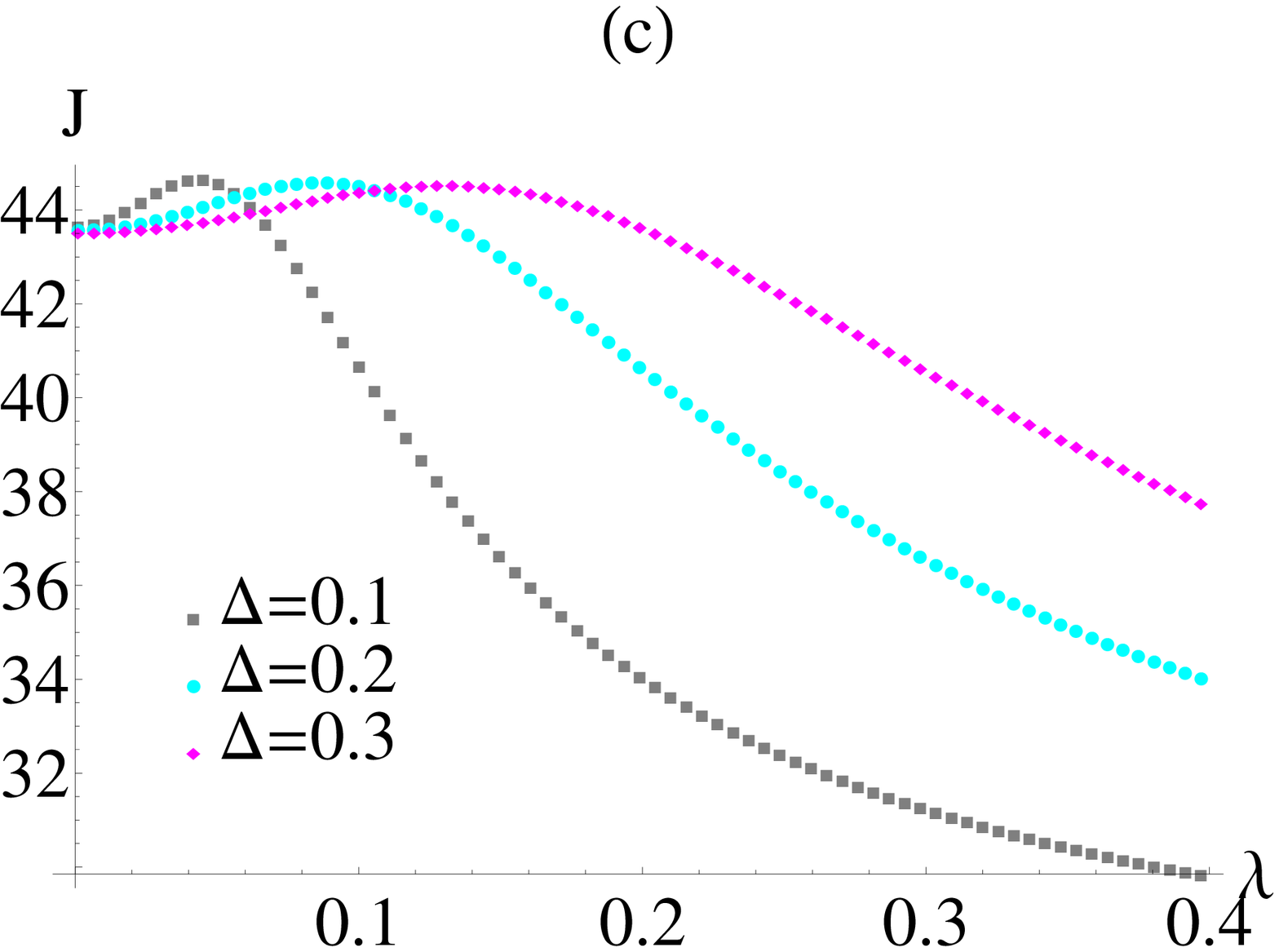}
\includegraphics[width=0.90\columnwidth]{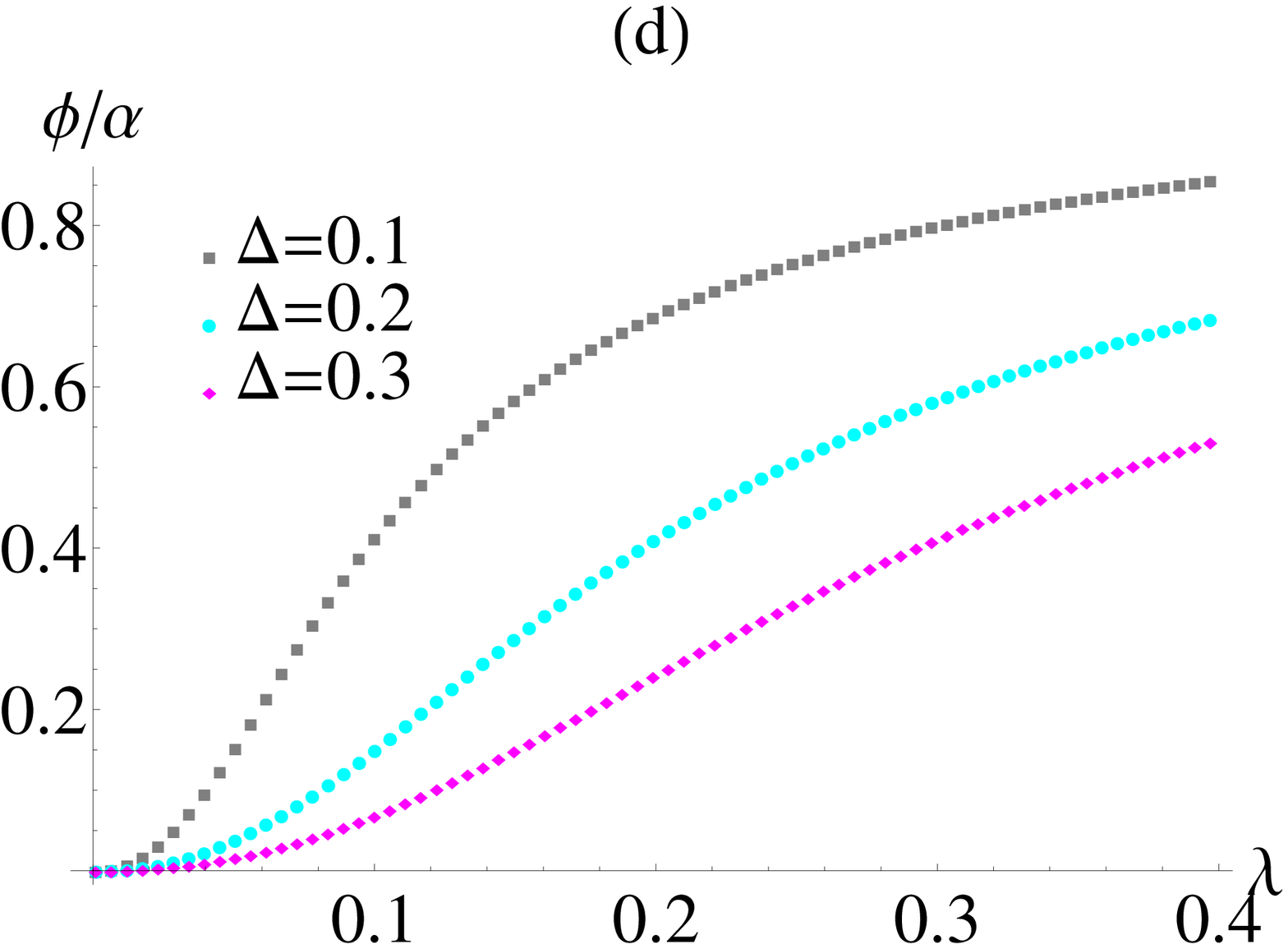}
\caption{
(a)
Anisotropic exchange couplings $\delta J_{xy},\delta J_{z}$ and the DM constant $D$ (in meV) as functions of SOC constant $\lambda$ shown by red diamond, green circle and blue square lines, respectively.
(b) The DM constant $D$ (in meV) as  function of  rotation angle $\alpha$. Black square, cyan circle, blue diamond lines correspond to  $\lambda=0.2;0.3;0.4$ eV, respectively.
(a) and (b)  The tetragonal field is considered to be equal to $\Delta=0.2$ eV.
(c) The isotropic exchange $J$ (in meV) and  (d) the  spin canting angle $\phi$ (in units of $\alpha$)
 as functions of SOC constant $\lambda$. Gray square, cyan circle, magenta diamond lines correspond to $\Delta=0.1;0.2;0.3$ eV, respectively.
 (a),(c) and (d) The rotation angle is considered to be equal to $\alpha=0.2$ rad.
The remaining parameters are
 $J_H=0.5$ eV, $U_2=2.5$ eV, and $t_{eff}=0.1$ eV. }
\end{figure*}

\subsection{ Application to Sr$_2$Ir$_{1-x}$Rh$_x$O$_4$.}

In this section we take a look at the properties of the Sr$_2$Ir$_{1-x}$Rh$_x$O$_4$ family of compounds, which results from substituting Sr$_2$IrO$_4$ by Rh ions.~\cite{qi12} Doping iridates with Rh ions does not change the band filling since Rh and Ir are in the same family of elements. However, the 4d orbitals of Rh are smaller than the 5d orbitals of Ir, which leads to a higher Coulomb repulsion, Hund's coupling and tetragonal CF splitting. Smaller atoms (such as Rh) also have reduced relativistic effects, including SO coupling. The effective hopping is reduced both due to a smaller overlap of the less extended 4d orbitals as well as due to the increased Coulomb repulsion. All of this  increases the importance of correlation effects and CF splitting as compared to that of SOC.

We also note that Sr$_2$RhO$_4$, the limiting case of Rh doping, shares with Sr$_2$IrO$_4$ the structural feature of staggered octahedra rotations about the axis perpendicular to the Rh/Ir planes. The angles of rotation are  similar to each other: $\sim 9.4^{\circ}$ for Sr$_2$RhO$_4$ and $\sim 12^{\circ}$ for Sr$_2$IrO$_4$.~\cite{qi12} This structure is preserved at intermediate levels of doping. As a result in the doped compounds the same interaction components are present as in pure Sr$_2$IrO$_4$ (Coulomb, Hund, SOC, CF, oxygen-assisted hopping, lattice distortions). Thus we expect the magnetic Hamiltonian to have the same structure of interactions. What changes is the overall energy balance of on-site interactions,   which in the Rh-doped iridates eventually leads to the appearance of different  magnetic phases compared to those in  pure iridium  compounds.

 Let us note that a reduced hopping simply leads to a decrease of the energy scale of all interactions (both the isotropic term and the anisotropies). The effects of reduced SOC are more intricate since SOC is the only interaction that mixes the orbital and the spin degrees of freedom of the holes. Thus we also expect the energy scale of the anisotropies to be reduced as compared to the isotropic term. This, however, does not diminish the importance of the anisotropic terms as their essential role is  to break the SU(2) symmetry of the isotropic interaction.

In Fig. 6 (a) and (c)  we plot the dependencies of the anisotropic and isotropic interactions, respectively, on the strength of $\lambda$. We  adjust all other parameters in accordance with the rhodium doping picture discussed above: $\Delta=0.2$ eV, $\alpha=0.2$ rad, $J_H=0.5$ eV, $U_2=2.5$ eV, and $t_{eff}=0.1$ eV.  A similar set of parameters might be realized at  low Rh doping. As expected the overall energy scale of the magnetic interactions is decreased due to Coulomb repulsion and smaller hopping. In Fig. 6 (a) we also see that once the SOC becomes too small relative to other interactions to effectively mix the orbital and spin degrees of freedom the anisotropic interactions quickly drop and reach zero in the limit of no SOC.

The behavior of the isotropic interaction is more interesting as it is weakly nonmonotonic as a function of SO-coupling. This can be explained when we look at different values of $\Delta$. Gray, cyan, and magenta lines in Fig. 6 (c) show the isotropic interaction corresponding to $\Delta=0.1,0.2,0.3$ eV, respectively. Since cubic orbitals have different hopping amplitudes due to the staggered rotations of the octahedra and the orbital symmetries (see Appendix B), the orbital composition of the ground state also determines the isotropic part of the exchange Hamiltonian. As the orbital composition is very sensitive to the interplay between the SOC and the CF, the small changes in their relative contribution  lead to a  non-monotonic  behavior of $J$.

 Fig. 6 (b) shows the dependence of the DM anisotropy on the staggered rotation angle $\alpha$ computed for $\lambda=0.2;0.3;0.4$ eV (respectively, black, cyan and  blue lines). As is the case for the pure Sr$_2$IrO$_4$ compound the DM interaction depends heavily on the angle but as we discussed above the overall range of DM interactions is smaller.

Finally, in Fig. 6(d) we present the  spin canting angle $\phi$ (in units of $\alpha$) as a function of the SOC constant $\lambda$ for various values of the tetragonal distortion $\Delta$. As expected, the canting angle is zero in the limit of zero SOC and is increasing with increasing $\lambda$, demonstrating  the key role of SOC in  the entanglement  of the spin and lattice  degrees of freedom.
 As we discussed above, the spin canting angle is  suppressed by the tetragonal distortion.

Our findings are in a qualitative agreement with experimental findings for the Sr$_2$Ir$_{1-x}$Rh$_x$O$_4$ family of compounds showing that Rh doping rapidly suppresses the magnetic transition temperature $T_C$ from 240 K at $x=0$ to almost zero at $x = 0.16$.\cite{qi12} The disappearance of long range magnetic order at small doping
in real compounds is a rather complicated phenomenon, related to both the reduction of magnetic interactions but also to the more metallic behavior of doped compounds.
 This aspect, however, can not be considered in our approach based on the assumption of a Mott insulator.
  We can only speculate that in Rh-doped iridates the
splitting between the ${\mathbf J}= 1/2$ and the ${\mathbf J}= 3/2$ states is substantially smaller  than in  pure iridium systems and, consequently, these two manifolds are  strongly mixed  by both Hund's coupling and tetragonal CF. The latter leads to a wider bandwidth and more metallic behavior. Interestingly, at high Rh doping, the system again becomes insulating, however for rather different reasons. There is an energy level mismatch for the Rh and Ir sites that makes the hopping of the carriers between  Rh and Ir ions more difficult.  The randomness of the Rh/Ir occupations gives rise
to Anderson localization and an insulating state.\cite{qi12} The magnetically ordered phase reappeared at $x>0.4$, but because of frustration it has rather low ordering temperature $T_C$ of the order  half a kelvin. This magnetic phase  needs to be studied in detail  both  experimentally and theoretically.

\section{Conclusion}\label{sec:conclusion}

In this paper we  provided  a theoretical framework   for the derivation of the
 effective super-exchange Hamiltonian governing magnetic properties of  transition metal oxides  with partially filled 4d and 5d shells.  We particularly focused on  iridates and rhodates -- materials which exhibit a rich variety of behavior owing to the interplay of correlation effects, strong SOC, and lattice distortions.
Our  approach allows one to relate the non-trivial magnetic behavior observed in these materials to their microscopic parameters. We show that the pseudospin super-exchange interactions governing the magnetic properties of this class of insulating materials  have anisotropic components of unusual types, leading not only to a dimensional reduction in pseudospin space (i.e., easy plane or easy axis anisotropy), but also to the chiral DM interaction and to additional frustration by bond-dependent interactions.

 How to derive exchange couplings from a given Hubbard-type Hamiltonian in the Mott-insulating regime is generally well understood. What gives rise to a certain complexity in our case is the combination of interactions and single particle energy shifts operating in different Hilbert subspaces. We restrict our consideration to the case of a ground state configuration of a single hole per transition metal ion in a pseudospin doublet state in one of the d-orbital multiplets. The exchange process then involves intermediate states with either zero holes or two holes. The latter states are governed by the Coulomb interaction components, especially the Hund's coupling. These ionic eigenstates need to be constructed and must be projected on to the single particle states describing the hopping processes. The resulting exchange couplings are then given by summation over all relevant intermediate eigenstates of the corresponding hopping element squared over the excitation energy of the  intermediate state.

The proposed theoretical approach can be  applied to compute exchange couplings  in iridium and rhodium oxides with different lattice geometries.  Many of these systems have been suggested as candidates for either interesting  magnetic orders or spin-liquid behavior in the Mott insulator regime.
Although the approach is quite general and can be applied to a variety of compounds, in this paper we focused on the  single-layer Sr$_2$IrO$_4$ and  Sr$_2$Ir$_{1-x}$Rh$_x$O$_4$ compounds, which have received much attention recently.
 For these systems we  first  derive the isotropic and anisotropic interactions analytically
and then study their dependencies on  microscopic parameters such as Hund's  and SOC coupling, and the strength of the lattice distortions.

Our results are the following. First, the overall strengths of the exchange couplings calculated by us appear to be in good agreement with experimental data, where available. While the  Ising-like and pseudospin anisotropic interactions are typically not larger than ten percent of the isotropic exchange, the  DM coupling is unusually large. It may be as large as 50\% of the isotropic exchange coupling for realistic values of the octahedra rotation angle. This emphasizes the importance
 of the SOC.

 We  computed the magnetic phase diagram of the model in the approximation of treating the pseudospins as classical objects..
We show that   for the parameter set most closely corresponding to the actual microscopic parameters of Sr$_2$IrO$_4$,
the magnetic ground  state of this compound is a coplanar canted antiferromagnet. This finding is in agreement with the experimental observation of the weak ferromagnetic moment accompanying the ground-state antiferromagnetic order in Sr$_2$IrO$_4$. We computed the spin canting angle and show that its  magnitude scales with the angle of the staggered rotations of the IrO$_6$ octahedra, as observed experimentally.

Finally, we studied how the properties of the  pure iridium systems are changed with Rh doping.  We show that Rh doping significantly modifies the hierarchy of many-body and single-particle interactions: the weaker SOC
combined with a stronger Coulomb interaction on Rh sites lead to overall smaller magnetic interactions and a weaker  coupling between  magnetic and structural degrees of freedom.

{\it Acknowledgement.}
We thank G. Jackeli and G.-W. Chern for useful discussions.
N.P. and Y.S. acknowledge the support from NSF grant DMR-1255544.
N.P. is also grateful for the hospitality of the visitors program at MPIPKS, where a part of the work on this manuscript has been done.  P.W. thanks the Department of Physics at the University of Wisconsin-Madison for hospitality during a stay as a visiting professor and acknowledges an ICAM senior scientist fellowship. P.W. also acknowledges partial support through the DFG research unit "Quantum phase transitions".

\appendix

\section{The structure of  the Hamiltonian matrix (14)}
\label{a0}

The block structure of the Hamiltonian matrix has the following form.
States $\left\vert {\mathcal J},1\right\rangle $ and $\left\vert {\mathcal J},10\right\rangle $
form the first block. The eigenstates are
\begin{eqnarray}\label{1_10}
\left\vert D,1\right\rangle &=&c_{1,1}\left\vert {\mathcal J},1\right\rangle
+c_{1,10}\left\vert {\mathcal J},10\right\rangle \\\nonumber
\left\vert D,10\right\rangle &=&c_{10,1}\left\vert {\mathcal J},1\right\rangle
+c_{10,10}\left\vert {\mathcal J},10\right\rangle
\end{eqnarray}
where $c_{1,1},\, c_{1,10},\, c_{10,1},\,c_{10,10}$ are the components of eigenvectors which  are obtained by diagonalizing (\ref{hamnottilde}).
We denote the eigenvalues of this block as
$E_1$ and $E_{10}$.

\bigskip

The second block involves states $\left\vert {\mathcal J},8\right\rangle$ and
$\left\vert {\mathcal J},15\right\rangle $. The Hamiltonian matrix for $(8,15)$ is identical to the one for $(1,10)$. Therefore,
\begin{eqnarray}\label{8_15}
\left\vert D,8\right\rangle &=&c_{8,8}\left\vert {\mathcal J},8\right\rangle
+c_{8,15}\left\vert {\mathcal J},15\right\rangle \\\nonumber
\left\vert D,15\right\rangle &=&c_{15,8}\left\vert {\mathcal J},8\right\rangle
+c_{15,15}\left\vert {\mathcal J},15\right\rangle \end{eqnarray}
with $c_{8,8}=c_{1,1}$, $c_{15,15}=c_{10,10}$, $c_{8,15}=c_{1,10}$ and $c_{15,8}=c_{10,1}$.
  The eigenenergies  of this block are simply
$E_{8}=E_{1}$ and $E_{15}=E_{10}$.

\bigskip

The third block involves the three states $\left\vert {\mathcal J},2\right\rangle
,\left\vert {\mathcal J},3\right\rangle ,\left\vert {\mathcal J},11\right\rangle $.  The eigenstates are given by
\begin{eqnarray}\label{2_3_11}
\left\vert D,2\right\rangle &=&c_{2,2}\left\vert {\mathcal J},2\right\rangle
+c_{2,3}\left\vert {\mathcal J},3\right\rangle +c_{2,11}\left\vert {\mathcal J},11\right\rangle\nonumber \\
\left\vert D,3\right\rangle &=&c_{3,2}\left\vert {\mathcal J},2\right\rangle
+c_{3,3}\left\vert {\mathcal J},3\right\rangle +c_{3,11}\left\vert {\mathcal J},11\right\rangle \\ \nonumber
\left\vert D,11\right\rangle &=&c_{11,2}\left\vert {\mathcal J},2\right\rangle
+c_{11,3}\left\vert {\mathcal J},3\right\rangle +c_{11,11}\left\vert {\mathcal J},11\right\rangle
\end{eqnarray}
and  eigenvalues  are $E_{2},E_{3},E_{11}$.

\bigskip

The fourth block  consists of (7,6,14) states  and is equivalent to the (2,3,11) block. The eigenstates are given by
\begin{eqnarray}\label{7_6_14}
\left\vert D,7\right\rangle &=&c_{7,7}\left\vert {\mathcal J},7\right\rangle
+c_{7,6}\left\vert {\mathcal J},6\right\rangle +c_{7,14}\left\vert {\mathcal J},14\right\rangle\nonumber \\
\left\vert D,6\right\rangle &=&c_{6,7}\left\vert {\mathcal J},7\right\rangle
+c_{6,6}\left\vert {\mathcal J},6\right\rangle +c_{6,14}\left\vert {\mathcal J},14\right\rangle \\ \nonumber
\left\vert D,14\right\rangle &=&c_{14,7}\left\vert {\mathcal J},7\right\rangle
+c_{14,6}\left\vert {\mathcal J},6\right\rangle +c_{14,14}\left\vert {\mathcal J},14\right\rangle
\end{eqnarray}

\bigskip

The fifth block  consists of (4,5,9,12,13) states. The eigenstates are given by
\begin{widetext}
\begin{eqnarray}\label{4_5_9_12_13}
\left\vert D,4\right\rangle &=&c_{4,4}\left\vert {\mathcal J},4\right\rangle
+c_{4,5}\left\vert{\mathcal J},5\right\rangle +c_{4,9}\left\vert {\mathcal J},9\right\rangle
+c_{4,12}\left\vert {\mathcal J},12\right\rangle +c_{4,13}\left\vert {\mathcal J},13\right\rangle
\nonumber \\
\left\vert D,5\right\rangle &=&c_{5,4}\left\vert {\mathcal J},4\right\rangle
+c_{5,5}\left\vert{\mathcal J},5\right\rangle +c_{5,9}\left\vert {\mathcal J},9\right\rangle
+c_{5,12}\left\vert {\mathcal J},12\right\rangle +c_{5,13}\left\vert {\mathcal J},13\right\rangle
\nonumber \\
\left\vert D,9\right\rangle &=&c_{9,4}\left\vert {\mathcal J},4\right\rangle
+c_{9,5}\left\vert {\mathcal J},5\right\rangle +c_{9,9}\left\vert {\mathcal J},9\right\rangle
+c_{9,12}\left\vert {\mathcal J},12\right\rangle +c_{9,13}\left\vert {\mathcal J},13\right\rangle
 \\
 \left\vert D,12\right\rangle &=&c_{12,4}\left\vert {\mathcal J},4\right\rangle
+c_{12,5}\left\vert {\mathcal J},5\right\rangle +c_{12,9}\left\vert {\mathcal J},9\right\rangle
+c_{12,12}\left\vert {\mathcal J},12\right\rangle +c_{12,13}\left\vert {\mathcal J},13\right\rangle
\nonumber \\\nonumber
\left\vert D,13\right\rangle &=&c_{13,4}\left\vert {\mathcal J},4\right\rangle
+c_{13,5}\left\vert {\mathcal J},5\right\rangle +c_{13,9}\left\vert {\mathcal J},9\right\rangle
+c_{13,12}\left\vert {\mathcal J},12\right\rangle +c_{13,13}\left\vert {\mathcal J},13\right\rangle
\end{eqnarray}
\end{widetext}
\bigskip
Note that the same block structure survives in the presence of the lattice distortions.

\section{Hopping operator for 180$^{\circ}$ Ir-O-Ir bond}

We first consider the  hopping operator  on the square lattice in the simplest case with no  tetragonal distortion, $\Delta =0$, and no rotations of the IrO$_6$ octahedra.
Without loss of generality we consider  $x$-bonds, and then using symmetry arguments  we  obtain transfer  matrix elements along  $y$-bonds.  Along an $x$-bond the hopping  occurs either
through  $p_{y}$- or $p_{z}$-orbitals of oxygen.
 The $p_{y}-$~orbital overlaps
with $\vert Z\rangle =\left\vert xy\right\rangle$  and $p_{z}-$~orbital
overlaps with $\left\vert Y\right\rangle =\left\vert zx\right\rangle $
orbitals of iridium.  Correspondingly, we denote the hopping amplitudes  as $t_{Z,y}$ and $t_{Y,z}$. However,   on the undistorted lattice $t_{Z,y}=t_{Y,z}$ and to simplify notations we denote the hopping amplitude as $t$.
Integrating over the oxygen ions, the effective hopping between Ir ions can be approximated as
$t_{eff}=t^2/\Delta_{pd}$, where $\Delta_{pd}$ is the charge-transfer gap. In our calculations we consider $t_{eff}=0.13$ eV.
The effective hopping Hamiltonian between Ir ions along the $x$-bond is then given by
\begin{eqnarray}
H_{t}^x=t_{eff}\sum_{n}\bigl(a_{Z\sigma ,n}^\dagger a_{Z\sigma
,n+x}+a_{Y\sigma ,n}^\dagger a_{Y\sigma ,n+x}+h.c.\bigr)
\end{eqnarray}
Expressing operators $a_{Z\sigma ,n}^\dagger$, etc.,  in terms of $b_{\gamma ,n}^\dagger$ operators, we get
\begin{eqnarray}
H_{t}^x=\sum_{n}\sum_{\gamma ,\gamma ^{\prime }}T_{n,n+x}^{\gamma ,\gamma
^{\prime }}(b_{n,\gamma }^\dagger b_{n+x,\gamma ^{\prime }}+h.c.)
\end{eqnarray}
where the  elements of the effective transfer matrix, $T_{n,n+x}^{\gamma ,\gamma ^{\prime }}$, between $\gamma$ and $\gamma^{\prime}$ orbitals  can be written as
\begin{eqnarray}
T_{n,n+x}^{\gamma ,\gamma ^{\prime }}=t_{eff}\left (\tau _{Z}^{\gamma ,\gamma
^{\prime }}+\tau _{Y}^{\gamma ,\gamma ^{\prime }}\right).
\end{eqnarray}
Here we use the following notation:
 \begin{eqnarray}\label{tauZandY} 
 \tau_Z ^{\gamma ,\gamma ^{\prime }} &=&\,_{\gamma^{\prime }}\langle J,J_z\vert{\hat T}\vert p^y_{\sigma}\rangle\langle p^y_{\sigma}\vert{\hat T}\vert J,J_z\rangle_{\gamma},\\\nonumber
 \tau_Y ^{\gamma ,\gamma ^{\prime }} &=&\,_{\gamma^{\prime }}\langle J,J_z\vert{\hat T}\vert p^z_{\sigma}\rangle\langle p^z_{\sigma}\vert{\hat T}\vert J,J_z\rangle_{\gamma}~,
 \end{eqnarray}

where ${\hat T}$ are hopping operators connecting neighboring Ir and O orbitals. In matrix form  $\tau_Z,\tau_Y$ are given by  

\begin{eqnarray}\label{tauz}
{\hat \tau}_{Z}=\frac{1}{3}\left(
\begin{array}{cccccc}
1 & 0 & 0 & -\sqrt{2} & 0 & 0 \\
0 & 1 & 0 & 0 & \sqrt{2} & 0 \\
0 & 0 & 0 & 0 & 0 & 0 \\
-\sqrt{2} & 0 & 0 & 2 & 0 & 0 \\
0 & \sqrt{2} & 0 & 0 & 2 & 0 \\
0 & 0 & 0 & 0 & 0 & 0%
\end{array}%
\right)
\end{eqnarray}

and

\begin{eqnarray}\label{tauy}
{\hat \tau}_{Y}=\frac{1}{6}\left(
\begin{array}{cccccc}
2 & 0 & 0 & \sqrt{2} & 0 & -\sqrt{6} \\
0 & 2 & \sqrt{6} & 0 & -\sqrt{2} & 0 \\
0 & \sqrt{6} & 3 & 0 & -\sqrt{3} & 0 \\
\sqrt{2} & 0 & 0 & 1 & 0 & -\sqrt{3} \\
0 & -\sqrt{2} & -\sqrt{3} & 0 & 1 & 0 \\
-\sqrt{6} & 0 & 0 & -\sqrt{3} & 0 & 3%
\end{array}%
\right)
\end{eqnarray}

In the presence of  tetragonal distortion and octahedra rotations, as in the case of Sr$_2$IrO$_4$,  the transfer matrix elements are more conveniently described using the global coordinate system.
The hopping between $\gamma$  and $\gamma ^{\prime }$ states is then given by
\begin{eqnarray}
{\bar T}_{n,n+x}^{\gamma ,\gamma ^{\prime }}=t_{eff}\left({\bar\tau}_{Z}^{\gamma ,\gamma
^{\prime }}+{\bar\tau}_{Y}^{\gamma ,\gamma ^{\prime }}\right),
\end{eqnarray}

where modified  transfer matrices are defined as

\begin{eqnarray}\label{tauZandYAB} 
 {\bar\tau}_Z ^{\gamma ,\gamma ^{\prime }} &=&\,_{\gamma^{\prime }}\langle \Psi_B\vert{\hat T}\vert p^y_{\sigma}\rangle\langle p^y_{\sigma}\vert{\hat T}\vert \Psi_A\rangle_{\gamma},\\\nonumber
 {\bar \tau}_Y ^{\gamma ,\gamma ^{\prime }} &=&\,_{\gamma^{\prime }}\langle \Psi_B\vert{\hat T}\vert p^z_{\sigma}\rangle\langle p^z_{\sigma}\vert{\hat T}\vert \Psi_A\rangle_{\gamma}.
 \end{eqnarray}

 Explicitly,  ${\hat {\bar\tau}}_{Z}$ and ${\hat {\bar\tau}}_{Y}$ are given by

\begin{eqnarray}\label{tauzAB}
{\hat {\bar\tau}}_{Z}=c^2_{\alpha}
\left(
\begin{array}{cccccc}
s^2_{\theta} & 0 & 0 & s_{\theta}c_{\theta} & 0 & 0 \\
0 & s^2_{\theta}& 0 & 0 & s_{\theta}c_{\theta} & 0 \\
0 & 0 & 0 & 0 & 0 & 0 \\
s_{\theta}c_{\theta}& 0 & 0 & c^2_{\theta} & 0 & 0 \\
0 & s_{\theta}c_{\theta}& 0 & 0 & c^2_{\theta} & 0 \\
0 & 0 & 0 & 0 & 0 & 0%
\end{array}%
\right)
\end{eqnarray}

and

\begin{eqnarray}\label{tauyAB} 
&&{\hat {\bar\tau}}_{Y}=\frac{1}{2}\\\nonumber&&\left(
\begin{array}{cccccc}
c^2_{\theta}e^{2\imath\alpha} & 0 & 0 & -s_{\theta}c_{\theta}& 0 & c_{\theta} e^{2\imath\alpha}\\
0 &  c^2_{\theta} e^{-2\imath\alpha} & -c_{\theta} e^{-2\imath\alpha} & 0&-s_{\theta}c_{\theta}& 0 \\
0 &-c_{\theta} e^{-2\imath\alpha}  &  e^{2\imath\alpha} & 0 &s_{\theta}  & 0 \\
-s_{\theta}c_{\theta} & 0 & 0 &s^2_{\theta}e^{2\imath\alpha}  & 0 & -s_{\theta} \\
0 & -s_{\theta}c_{\theta} &s_{\theta}  & 0 & s^2_{\theta}e^{-2\imath\alpha} & 0 \\
c_{\theta} e^{-2\imath\alpha} & 0 & 0 & -s_{\theta} & 0 & e^{-2\imath\alpha}%
\end{array}%
\right)
\end{eqnarray}

\end{document}